\documentclass{article}[11pt]
\usepackage{amsmath, amsfonts,amssymb,amsthm}
\usepackage[dvips]{graphicx}
\usepackage{cases,color,natbib,setspace}
\usepackage{mathtools}
\usepackage{hyperref}
\usepackage{subfig}
\hypersetup{citecolor=green}
\usepackage[margin=1in]{geometry}

\newtheorem{theorem}{Theorem}[section]
\newtheorem{exa}{Example}[section]
\newtheorem{corollary}{Corollary}[section]
\newtheorem{lemma}{Lemma}[section]
\newtheorem{prop}{Proposition}[section]

\newtheorem{Assumption}{Assumption}[section]
\newtheorem{remark}{Remark}[section]

\newcommand{\be}{\begin{equation}}
\newcommand{\ee}{\end{equation}}

\newcommand{\bq}{\begin{eqnarray}}
\newcommand{\eq}{\end{eqnarray}}
\newcommand{\ex}{\mathbb{E}}
\newcommand{\half}{\frac{1}{2}}

\newcommand{\ind}{\mathbf{1}} 
\newcommand{\nn}{\nonumber}

\newcommand{\sk}{\smallskip}
\newcommand{\ul}{\underline}
\newcommand{\ol}{\overline}
\newcommand{\mc}{\mathcal}

\newcommand{\e}{\mathbf{e}}

\newcommand{\pr}{\mathbb{P}}

\newcommand{\diff}{\textup{d}}
\newcommand{\R}{\mathbb{R}}
\newcommand{\et}{\mathrm{e}}
\newcommand{\es}{\epsilon}

\def\1{{\mathbf 1}}

\begin{document}


\title{\Large \textbf{Optimal trading with a trailing stop}}
\author{Tim Leung\thanks{ \small{Department of Applied Mathematics, University of Washington, Seattle WA 98195. E-mail:
\mbox{timleung@uw.edu}.  Corresponding author. }} \and Hongzhong Zhang\thanks{IEOR Department, Columbia University. New York, NY 10027. E-mail: \mbox{hz2244@columbia.edu}.  }}
\maketitle
\begin{abstract}
 Trailing stop is a popular stop-loss trading strategy by which the investor will sell the asset once its price experiences a pre-specified percentage drawdown. In this paper, we study the problem of timing to  buy and then sell an asset subject to a trailing stop. Under   a general linear diffusion framework, we study an optimal double stopping problem with a random path-dependent maturity. Specifically, we first analytically solve the optimal liquidation problem with a trailing stop, and in turn derive the optimal timing to  buy the asset.  Our method of solution reduces the problem of determining the optimal trading regions to solving the associated differential equations. For illustration, we  implement  an   example and conduct a sensitivity analysis under  the exponential Ornstein-Uhlenbeck model. 
\end{abstract}

\vspace{10pt}
 
 

\newpage
\onehalfspacing

\section{Introduction}
Trailing stops are a popular trade order widely used by proprietary traders and  retail investors   to provide  downside protection for  an existing position. In contrast to a stop-loss exit that closes a position at a fixed price, a  trailing stop is characterized by  a stochastic floor that moves based on the  running maximum of the asset price.  This provides a dynamic  downside protection as the stochastic floor is automatically adjusted upward whenever the asset price moves to a new high. A trailing stop is triggered when the prevailing price of an asset falls below the stochastic floor.  In essence, it allows  an investor to specify a limit on the maximum possible loss  while not limiting the maximum possible gain. This  is particularly relevant in common trend-following strategies, and trailing stop provides an automatic trigger to exit when prices start to trend downward due to, for example, regime switching \citep{DaiTrend2010}.

In addition to setting a trailing stop order, the investor can also use a limit order to sell at a certain price target. If the price is sufficiently high, the investor may prefer to take profit immediately, rather than waiting further with the possibility of setting off the trailing stop. The investor's position will be liquidated by either order, whichever comes first.  

In this paper, we investigate the mathematical problem of optimal timing to liquidate a position subject to a trailing stop.  Mathematically, we recognize the trailing stop as a \emph{stochastic  timing constraint} in the sense that it installs a path-dependent random  maturity into the  liquidation problem, rending the problem significantly more difficult to analyze or solve. Furthermore, the investor can decide when to establish the position in the first place. This leads us to also analyze the optimal timing to enter the market. In sum, we study an optimal double stopping problem subject to a trailing stop. Using excursion theory of linear diffusion, we derive the value functions using  the smallest concave majorant characterization, and discuss the effect of trailing stopping on the optimal trading  strategies analytically and numerically. Among our results, we reduce the problem of finding the optimal timing strategies to solving an ODE problem, which forms the basis of our numerical scheme in determining the optimal asset acquisition and liquidation regions. 
 
%
%

In general, a trailing stop can be defined as the first time when the asset price $X$ drops below $f(\ol{X})$, where $\ol{X}$ is the running maximum process of $X$, and $f$ is an increasing function such that $f(x)<x$ for all $x$ in the support of $X$.
In applied probability literature, such a stopping time is related to the drawdown process and its first passage time. We refer to   \cite{Lehoczky77},   \cite{ZhangAAP15}, and   \cite{ZhanHadj12}, for  a partial list of studies on drawdowns under linear diffusions. Moreover, the optimality of trailing stops in exercising (generalized) Russian options and detecting abrupt changes can be found in   \cite{shepp-shiryaev}, \cite{Egami_Oryu_FS17} and  \cite{Zhang2015}, respectively

Despite being commonly used by practitioners, 
  trailing stops have been  scarcely studied  in the mathematical finance literature.   We trace back to  \cite{Glynn1995}, who studied the expected discounted reward at a trailing stop under a discrete-time random walk or a geometric Brownian motion (GBM) model, and found that it would be optimal to never use the trailing stop if the stock followed a GBM with  a positive drift. In contrast, our study is conducted in a more general linear diffusion framework, and provides concrete illustrative example on how to the use of a trailing stop will affect the optimal timing to sell an asset under the exponential Ornstein-Uhlenbeck model.  In a random walk model, \cite{Warburton2006} performed a probabilistic analysis of a variant of trailing stop.  \cite{Yin2010} implemented a stochastic approximation scheme to determine the optimal percentage trailing stop level that maximizes the expected discounted simple return from  liquidation. The   recent study by \cite{Imkeller2014} compared   the performance of a number of  trading rules with fixed and trailing stops under an arithmetic  Brownian motion model.

Compared to these works, we tackle the trading problem by formulating an  \emph{optimal double stopping} with a \emph{stochastic timing constraint} induced by the trailing stop, and we rigorously derive the optimal trading strategy. Our method of solution applies to a general linear diffusion framework, and our analytical results   facilitate  computation of the value function and optimal timing strategies (see Section \ref{sec:xou}).   In our optimal liquidation problem subject to the trailing stop, we show that it is optimal to use a limit sell order at a sufficiently high price. In other words, once  the investor enters the market, he/she can immediately set the optimal limit sell order together with the trailing stop order, and wait for either order to be executed automatically.  

The  trailing stop can be viewed as a random maturity or stopping time constraint in the optimal stopping problem, in the sense that any admissible stopping time must come before  triggering the  trailing stop. Related studies by the authors include  optimal stopping problems with maturities determined by an occupation time  (\cite{OmegaRZ15,Rodosthenous2017}) or by a default time (\cite{Leung2013}), and optimal mean reversion trading with a fixed   stop-loss exit (\cite{LiLeung15}). In particular, part of our study (Section \ref{opt1}) generalizes the analytical framework of \cite{LiLeung15} to general linear diffusions, and the results from optimal stopping subject to  a fixed stop-loss exit will prove to be directly  useful for  solving  the analogous problem with a trailing stop.

%
%
%
%

The remaining of the paper is structured as follows.   Section \ref{sec:MF} presents   stochastic framework for our trading problem. In  Section \ref{opt1}, we study an  optimal trading problem with a fixed stop-loss. Then, in Section \ref{sec:trailing}, we study the  optimal stopping problems for  trading with a trailing stop. To illustrate  our analytical results, we consider trading under the  exponential Ornstein-Uhlenbeck model, and   numerically compute  the optimal acquisition  and liquidation regions  in Section \ref{sec:xou}.  We also provide a sensitivity analysis on  the  optimal trading strategies  with respect to model parameters.  Detailed proofs  are collected in the Appendix.

\section{Model Formulation}\label{sec:MF}
Let us consider  a risky asset value process $X_\cdot=\{X_t\}_{t\ge0}$ modeled by  a  linear diffusion on $I\equiv(l,r)\subset\R$ with the infinitesimal generator: 
\be \label{generator}
\mc{L}=\frac{1}{2}\sigma^2(x)\frac{\partial^2}{\partial x^2}+\mu(x)\frac{\partial}{\partial x},\quad\forall x\in I,
\ee
where $(\mu(\cdot),\sigma(\cdot))$ is a pair of real-valued functions on $I$ such that 
\[\frac{1+|\mu(\cdot)|}{\sigma^2(\cdot)}\in\mathbb{L}^1_{\textup{Loc}}(I)\quad \text{ and }\quad \sigma(x)>0,\,\quad \forall x\in I.\]
For any $\bar{x}\in I$, the running maximum of $X$ is denoted by \[\ol{X}_t:=\bar{x}\vee\sup_{s\in[0,t]}X_s,\qquad t\ge 0.\]  

We denote the unique  probability law of  $X_\cdot$ by   $\pr_{x,\bar{x}}$ given $\{X_0=x,\ol{X}_0=\bar{x}\}$ for any $x,\bar{x}\in I$ with $x\le\bar{x}$. The   expectation associated with   $\pr_{x,\bar{x}}$ is denoted by  $\ex_{x,\bar{x}}$. In calculations and results where   the initial value $\ol{X}_0=\bar{x}$ is irrelevant, we simply  write  $\pr_{x}$ and $\ex_{x}$ to denote the probability law of $X_\cdot$ and the associated  expectation given $\{X_0=x\}$. Throughout, we assume that the both boundaries $l,r$ are inaccessible. 

\begin{figure}
\centering
\includegraphics[width=3.5in]{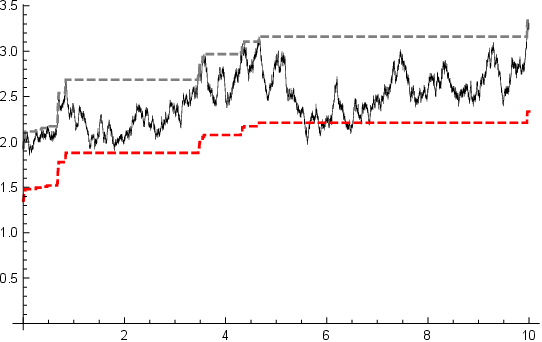}
\caption{Sample paths of the asset price (solid black), its running maximum (gray dashed), and the $30\%$-drawdown floor  representing the trailing stop (red dashed). }\label{fig0}
\end{figure}

We consider an investor who holds long one unit of the risky asset $X$.  Our objective is to investigate the optimal trading strategy with a trailing stop. To this end, we consider the problem of optimal early liquidation of this risky asset, given a pre-specified trailing stop mandatory liquidation order. Specifically, we will model liquidation time by a stopping time $\tau$ of the underlying process $X_\cdot$, and the   reward to be realized upon  liquidation by $h(X_\tau)$, where    
 $h(\cdot)$ is a real-valued increasing function on $I$, such that $\{x\in I:h(x)>0\}\neq\emptyset$. Fix a function $f(\cdot)$ on $I$, such that 
 \be
 \begin{array}{cc}f(\cdot)\text{ is continuous, strictly increasing on }I,\\ \text{for all }x\in I, f(x)\in I, f(x)<x. \end{array}\label{eq:floor}\ee
Then, we define the \emph{stochastic floor} by  $f(\ol{X})$, where $\ol{X}$ is the running maximum  of $X$.  The trailing stop, denoted by $\rho_f$, is defined as the first time   the asset value $X$ reaches the  stochastic floor  $f(\ol{X})$ from above.  That is,\footnote{As usual, we set $\inf\emptyset=\infty$.}
 \be
 \rho_f:=\inf\{t>0: X_t<f(\ol{X}_t)\}. 
 \ee
 \begin{remark}\label{exam}
We give two  standard choices of the floor function $f(\cdot)$ here. For example,  if $I=\mathbb{R}$, setting $f(x)=x-a$ for some $a>0$ gives the absolute drawdown floor, and $\rho_f$ is the first time   $X$ falls  from its maximum $\ol{X}$ by $a$ units. Another specification when $I=\mathbb{R}_+$, $f(x)=(1-\alpha)x$ for some $\alpha\in(0,1)$, gives the percentage drawdown, and $\rho_f$ is the first time   $X$ falls  from its maximum $\ol{X}$ by $(100\times\alpha)\%$, as depicted in Figure \ref{fig0} with $\alpha=0.3$.
\end{remark}

  The  investor faces the following optimal stopping problem:
\be
v_f(x,\bar{x}):=\sup_{\tau\in\mc{T}_f^{\sf T}}\ex_{x,\bar{x}}(\et^{-q\tau}h(X_\tau)\ind_{\{\tau<\infty\}}).\label{eq:trailingproblem}
\ee
where $q>0$ is a subjective discounting rate, and $\mc{T}_f^{\sf T}$ is the set of all stopping times of $X$ that stop no later than the trailing stop $\rho_f$. Notice that $\rho_f$ puts a mandatory selling order of the risky asset, pre-specified  by the investor.

To quantify the gain  in terms of  expected discounted reward from  liquidating  earlier than the trailing stop  time $\rho_f$, we define  the  \emph{early liquidation premium} by the difference
\begin{align}
p_f(x,\bar{x}):=&v_f(x,\bar{x})-g_f(x,\bar{x}),\label{eq:pa}
\end{align}
where the second term represents the expected discounted reward from waiting to sell  at the trailing stop, that is, 
\be\label{eq:ga}
g_f(x,\bar{x}):=\ex_{x,\bar{x}}(\et^{-q\rho_f}h(X_{\rho_f})\ind_{\{\rho_f<\infty\}}).
\ee
As a convention, we define $\infty-\infty=\infty$ if both terms on the right-hand side of \eqref{eq:pa} are infinity. Clearly, we have $p_f(x,\bar{x})\ge0$ for all $x,\bar{x}\in I$ with $x\le \bar{x}$. For our study, the  early liquidation premium turns out to be amenable to analysis and give intuitive interpretations.  The related concepts of early/delayed exercise/purchase premium   have been analyzed in pricing American options (see \cite{Carr1992}) and derivatives trading (\cite{Leung2011b}), among other applications.

\begin{remark}\label{rem21}
If floor functions $f_1(\cdot), f_2(\cdot)$ both satisfy \eqref{eq:floor}, and $f_1(x)\le f_2(x)$ for all $x\in I$, then for every fixed $x\in I$, we have the inequalities: 
\be\label{eq:opt_wo_trailing} h(x) \le v_{f_2}(x,\bar{x})\le v_{f_1}(x,\bar{x})\le\sup_{\tau\in\mc{T}}\ex_{x}(\et^{-q\tau}h(X_\tau)\ind_{\{\tau<\infty\}}),\ee
where $\mc{T}$ is the set of all stopping times of $X$.
\end{remark}

Given the optimal value $v_f(x,\bar{x})$, another related  problem is
\be
\label{eq:trailingproblem2} v_f^{(1)}(x)=\sup_{\tau\in\mc{T}}\ex_{x}(\et^{-q\tau}(v_f(X_\tau,X_\tau)-h_b(X_\tau))\ind_{\{\tau<\infty\}}),
\ee
where  $h_b(\cdot)$ is an increasing function on $I$ such that $h_b(x)\ge h(x)$ for all $x\in I$,\footnote{If there is an $x\in I$ such that $h_b(x)<h(x)$, then immediate selling after purchasing when the asset price is at $x$ yields a strictly positive profit with certainty, hence an arbitrage.} and $\sup_{x\in I}(v_f(x,x)-h_b(x))>0$, $\mc{T}$ is the set of all stopping times w.r.t. the filtration generated by $X$. The problem arises, for example, in optimal acquisition of the asset $X$ when $h_b(x)=x+c_b$, where $c_b\ge0$ is a transaction fee.\footnote{In this case, $h(x)=x-c_s$ where $c_s\ge0$ is a transaction fee.} In general, if we assume that $h_b(X)$ is the price the investor need to pay to acquire  one unit of the risky asset, then \eqref{eq:trailingproblem2} represents the problem of finding  the optimal time to purchase this risk asset. Note that the investor will select the optimal time to sell but subject to a trailing stop exit. 
For this reason, we will call the problem in \eqref{eq:trailingproblem2}  the optimal acquisition problem with a trailing stop, even for a general reward $h_b(\cdot)$. 

\begin{remark}\label{rmk:practical}
Note that in \eqref{eq:trailingproblem2}, we apply the value function $v_f(x,\bar{x})$ \emph{only} with $x=\bar{x}$. From a practical point of view, this is the most relevant case since a trailing stop should be placed based on   the price at which the asset was purchased, rather than an arbitrary reference price. 
\end{remark}

In summary, the solutions to \eqref{eq:trailingproblem2} and \eqref{eq:trailingproblem} yield the optimal trading strategy that involves buying    a  risky asset and selling  it later while being protected by a trailing stop.

\subsection{Preliminaries of Linear Diffusions}

It is well known that (see, for example, \cite[]{Borodin2002}), for any $q>0$, the {\em Sturm-Liouville equation} $(\mathcal{L}-q)u(x)=0$ has a positive increasing solution $\phi_q^+(\cdot)$  and a positive decreasing solution $\phi_q^-(\cdot)$. In fact, for an arbitrary fixed $\kappa\in I$, the solutions can be expressed as
\be\label{eq:phi}
{\phi_q^+(x)=}
\begin{dcases}\ex_x(\textup{e}^{-q\tau_X^+(\kappa)}), &\text{if }x\le \kappa\\
\frac{1}{\ex_\kappa(\textup{e}^{-q\tau_X^+(x)})}, &\text{if }x>\kappa
\end{dcases},\,\,
{\phi_q^-(x)=}\begin{dcases}
\frac{1}{\ex_\kappa(\textup{e}^{-q\tau_X^-(x)})},&\text{if }x\le \kappa\\
{\ex_x(\textup{e}^{-q\tau_X^-(\kappa)})},&\text{if }x>\kappa
\end{dcases},
\ee
where $\tau_X^+(y)$ and  $\tau_X^-(y)$ are the first passage times of $X$ to level $y\in I$ from below and above, respectively, 
\be
\tau_X^\pm(y):=\inf\{t>0: X_t\gtrless y\},\quad\forall y\in I.
\ee

The functions $\phi_q^\pm(\cdot)$ are also closely related to two-sided exit problems of $X$. Specifically,  we have:
 \begin{lemma}\label{fundamentallemma}\cite{Lehoczky77}
Suppose that $l<y\le x\le z<r$, then for $q>0$, we have
\begin{align}
  \ex_x(\textup{e}^{-q\tau_X^-(y)}\ind_{\{\tau_X^-(y)<\tau_X^+(z)\}})&=\frac{\phi_q^-(x)}{\phi_q^-(y)}\frac{\psi_q(z)-\psi_q(x)}{\psi_q(z)-\psi_q(y)},\nn\\
  \ex_x(\textup{e}^{-q\tau_X^+(z)}\ind_{\{\tau_X^-(y)>\tau_X^+(z)\}})&=\frac{\phi_q^-(x)}{\phi_q^-(z)}\frac{\psi_q(x)-\psi_q(y)}{\psi_q(z)-\psi_q(y)},\nn
\end{align}
where $\psi_q:I\mapsto\R_+$ is a strictly increasing function defined as 
\be
\psi_q(x):=\frac{\phi_q^+(x)}{\phi_q^-(x)},\quad\forall x\in I.
\ee
\end{lemma}
\begin{remark}
By the boundary behavior of $X$, we have $\phi_q^-(l+)=\phi_q^+(r-)=\infty$, hence $\psi_q(I)=(0,\infty)$. See \cite[p. 18-19]{Borodin2002} for more details. 
\end{remark}
\subsection{Standing Assumption and its Implications}
We now discuss the following standing assumption on the reward function $h(\cdot)$.
\begin{Assumption}\label{assume1}
The reward function $h(\cdot)$ is increasing, twice differentiable on $I$, and there is an $x_0$ in the interior of  $I$ such that 
\be
(\mathcal{L}-q)h(x)\ge0\text{ if and only if }x\le x_0.\label{eq:trailing:assume1}\ee
Moreover, we have 
\be
\lim_{x\to r}\frac{h(x)}{\phi_q^+(x)}<\sup_{x>x_0}\frac{h(x)}{\phi_q^+(x)}<\infty,\label{eq:H0H}\ee
\end{Assumption}
\begin{remark}
By \cite[Proposition 5.10]{DK2003}, it is easily seen that Assumption \ref{assume1} ensures the finiteness of the upper bound in \eqref{eq:opt_wo_trailing}  for all $x\in I$. Moreover, the assumption implies that the optimal stopping time for the upper bound is of threshold type, as proved in the following lemma.\end{remark}
\begin{lemma}
\label{lem:threshold_opt}
Under Assumption \ref{assume1}, there is an $x^\star\in[x_0,r)$ such that, 
\[\sup_{\tau\in\mc{T}}\ex_x(\et^{-r\tau}h(X_\tau)\ind_{\{\tau<\infty\}})=\ex_x(\et^{-r\tau_X^+(x^\star)}h(X_{\tau_X^+(x^\star)})\ind_{\{\tau_X^+(x^\star)<\infty\}}).\]
\end{lemma}

Lemma \ref{lem:threshold_opt} shows that Assumption \ref{assume1} is sufficient for the optimality of upcrossing strategy $T_{x^\star}^+$ in optimal stopping problem $\sup_{\tau\in\mc{T}}\ex_x(\et^{-r\tau}h(X_\tau)\ind_{\{\tau<\infty\}})$, where $x^\star$ is a constant in $(x_0,r)$. Since   $h(X_\tau)$ represents  the proceeds from selling the risky asset,  the economic insight of Lemma \ref{lem:threshold_opt} is that, under no constraint (i.e. no trailing stop), it is optimal to sell the asset when its price is sufficiently high. Thus, apart from analytical tractability considerations, Assumption \ref{assume1} is also economically reasonable for our trading problem.

\begin{remark}
We give a few examples in which Assumption \ref{assume1} holds. First, let $h(x)=x-K$ for some constant $K>0$, and $X_\cdot$ be the Black-Scholes model, i.e. $\mu(x)=\mu x, \sigma(x)=\sigma x$ for all $x\in I=\mathbb{R}_+$, with constants $\mu<q, \sigma>0$.\footnote{It is well-known that if $\mu\ge q$, then the optimal stopping region is the empty set. } Second, we can let $h(x)=x$ and $X_\cdot$ be the Ornstein-Uhlenbeck process, i.e. $\mu(x)=\lambda(\theta-x)$ and $\sigma(x)=\sigma$ for $x\in I=\R$, with constants $\lambda,\sigma>0$ and $\theta\in\R$.
\end{remark}

\section{Optimal Trading with  a Fixed Stop-Loss}\label{opt1}
To gain some intuition for  our solution method for the problem in \eqref{eq:trailingproblem} 
with a trailing stop, we first consider the  optimal stopping problems when the investor  uses  a fixed stop-loss exit instead of a trailing stop. Precisely,  arbitrarily fix a $y\in I$, we consider the following class of problems indexed by $y$:
\begin{align}
V_y(x):=&\sup_{\tau\in\mc{T}_y^{\sf S}}\ex_{x}(\et^{-q\tau}h(X_\tau)\ind_{\{\tau<\infty\}}),\label{eq:fixedproblem}
\end{align}
where $\mc{T}_y^{\sf S}$ is the set of all stopping times of $X$ that stops no later than the first passage time to level $y$, i.e. \begin{align}
\tau_X^-(y)=\inf\{t>0: X_t<y\},\label{tauxy}\end{align} 
 and $c\in[0,\sup_{x\in I}(V_y(x)-h(x)))$ is a transaction fee for asset acquisition.
The problem in \eqref{eq:fixedproblem} puts a mandatory liquidation constraint upon hitting the fixed stop-loss level $y$ from above.

The special cases of  the  problem in   \eqref{eq:fixedproblem} 
with the reward function $h(x)=x-c$  driven by the OU and CIR processes have been studied  in \cite{Cartea2015,LiLeung15,Leung2014,LiLeungWang15}. In this section, we present the analysis of problem   \eqref{eq:fixedproblem} 
driven by a general linear diffusion.

\subsection{Optimal Liquidation Subject to   a Stop-loss Exit}
We now study  the optimal liquidation problem \eqref{eq:fixedproblem} where $X$ follows a general linear diffusion (see \eqref{generator}). To facilitate our analysis, we also consider the extended case of \eqref{eq:fixedproblem} for $y=l$, in which case we have 
\begin{align}\label{Vlx}V_l(x)=\sup_{\tau\in\mc{T}}\ex_x(\et^{-q\tau}h(X_\tau)\ind_{\{\tau<\infty\}}).\end{align}
Notice that the value function $V_l(\cdot)$ has already been derived in Lemma \ref{lem:threshold_opt}.
\begin{remark}\label{rem:mono}
For each fixed $x\in I$, the mapping $y\mapsto V_y(x)$ is obviously non-increasing over $[l,r)$.
\end{remark}
\begin{remark}\label{rem:32_trailing}
The connection between \eqref{eq:trailingproblem} and \eqref{eq:fixedproblem} can be seen as follows.
For any $x,\bar{x}\in I$ such that $x\in(f(\bar{x}),\bar{x}]$, by the $\pr_{x,\bar{x}}$-a.s. inequality that $\rho_f\le \tau_X^-(f(\bar{x}))$, we know that $\mc{T}_f^{\sf T}\subset\mc{T}_{f(\bar{x})}^{\sf S}$. Hence, $v_f(x,\bar{x})\le V_{f(\bar{x})}(x)$. As a consequence, if we define the optimal liquidation regions
\begin{alignat}{2}
\mc{S}_f^{\sf T, L}(\bar{x}):=&\{x\in(l,\bar{x}]: v_f(x,\bar{x})=h(x)\},\quad&&\forall \bar{x}\in I,\\
\mc{S}_{y}^{\sf S, L}:=&\{x\in I: V_y(x)=h(x)\},\quad&&\forall y\in I,
\end{alignat}
then we have \[\left(\mc{S}_{f(\bar{x})}^{\sf S, L}\cap(l,\bar{x}]\right)\subset\mc{S}_f^{\sf T, L}(\bar{x}), \qquad \forall  \bar{x}>0. \] Additionally, if $\bar{x}\in\mc{S}_{f(\bar{x})}^{\sf S, L}$ then we have $\left(\mc{S}_{f(\bar{x})}^{\sf S, L}\cap(l,\bar{x}]\right)=\mc{S}_f^{\sf T, L}(\bar{x})$, since in this case it is optimal to liquidate before $X$ reaching  a new maximum. 
\end{remark}

\begin{prop}\label{trailing:thm33}
Under Assumption \ref{assume1}, for any fixed $y\in(l,x_0)$, there is a finite threshold  $b(y)\in(x_0, r)$ 
such that \footnote{Notice that in the   expectation \eqref{vyx} we don't have the indicator $\ind_{\{\tau_X^+(b(y))\wedge\tau_X^-(y)<\infty\}}$, as it is equal to 1 almost surely.}  
\begin{align}\label{vyx}V_y(x)=\ex_x(\et^{-q(\tau_X^+(b(y))\wedge\tau_X^-(y))}h(X_{\tau_X^+(b(y))\wedge\tau_X^-(y)})),\quad\forall x\in I.\end{align}
Here $b(y)$ can be identified as the smallest solution over $(x_0,r)$ to 
\be h'(b)-h(b)\frac{\phi_q^{-,\prime}(b)}{\phi_q^-(b)}=\frac{\phi_q^-(b)\psi_q'(b)}{\psi_q(b)-\psi_q(y)}\left( \frac{h(b)}{\phi_q^-(b)}-\frac{h(y)}{\phi_q^-(y)}\right).\label{trailing:eq:8.20}\ee
Moreover, the mapping $y:\mapsto b(y)$ is strictly decreasing and differentiable over $(l,x_0)$, with limits $b(x_0-)=x_0$, and $b(l+)\le x_*<r$, where $x_*$ is defined in Lemma \ref{lem:threshold_opt}. \end{prop}

\begin{corollary}\label{cor34}
If $y\in[x_0,r)$, then the stopping region $\mc{S}_y^{\sf S, L}=I$, i.e. there is no continuation region. 
\end{corollary}

\section{Optimal Trading with  a Trailing Stop}\label{sec:trailing}
In this section, we apply the results we obtained to study the optimal liquidation problem \eqref{eq:trailingproblem} and the optimal acquisition problem \eqref{eq:trailingproblem2}.

\subsection{Optimal Liquidation  }
Returning to the problem in \eqref{eq:trailingproblem}, we will first using results in Theorem \ref{trailing:thm33} to construct a candidate threshold type strategy for liquidation before the trailing stop $\rho_f$.
\begin{corollary}\label{trailing:cor35}
There is a unique $b_f^\star\ge x_0$ such that $b(f(\bar{x}))>\bar{x}$ if and only if $\bar{x}<b_f^\star$.  Moreover, $b_f^\star$ can be identified as  the unique solution over $(x_0,f^{-1}(x_0))$ to $\Gamma(\bar{x})=0$, where
\be\Gamma(\bar{x}):=\frac{1}{\psi_q'(\bar{x})}\bigg(\frac{h'(\bar{x})}{\phi_q^-(\bar{x})}-\frac{h(\bar{x})\phi_q^{-,\prime}(\bar{x})}{(\phi_q^-(\bar{x}))^2}\bigg)-\frac{1}{\psi_q(\bar{x})-\psi_q(f(\bar{x}))}\left(\frac{h(\bar{x})}{\phi_q^-(\bar{x})}-\frac{h(f(\bar{x}))}{\phi_q^-(f(\bar{x}))}\right).\label{eq:cor42}\ee
Moreover, $\Gamma(\bar{x})>0$ if $l<\bar{x}<b_f^\star$, and $\Gamma(\bar{x})<0$  if $f^{-1}(x_0)>\bar{x}>b_f^\star$.
\end{corollary}
\sk
\begin{remark}
We briefly explain the rationale behind the characterization of $b_f^\star$ in Corollary \ref{trailing:cor35} here. Instead of solving the stopping problem with a trailing stop \eqref{eq:trailingproblem} optimally, let us consider a sub-optimal, myopic strategy. The strategy that will be considered is the optimal strategy given in Theorem \ref{trailing:thm33} when $y=f(\ol{x})$, where $\ol{x}$ is the initial level of the running maximum. We call this strategy myopic because the strategy is obtained by fixing a stop-loss level, instead of allowing the stop level moving along with the running maximum. Typically, it is impossible to attain the optimal myopic stopping threshold without establishing a new high for the running maximum, unless the initial maximum level is already sufficiently high.  The threshold $b_f^\star$ is the critical level beyond which, the above myopic strategy becomes optimal. In fact, when the initial running maximum $\ol{x}=b_f^\star$, the optimal stopping threshold for the fixed stop-loss level at $y=f(b_f^\star)$ is exactly at $b_f^\star$. Hence, \eqref{eq:cor42} is obtained by imposing  \eqref{trailing:eq:8.20} to hold when $b=\ol{x}$ and $y=f(\ol{x})$.
\end{remark}

Let us suppose for now  that $\bar{x}\ge b_f^\star$. Then,
\begin{enumerate}
\item If we still have $f(\bar{x})<x_0$, then by the definition of $b_f^\star$ given in Corollary \ref{trailing:cor35}, we have $b(f(\bar{x}))\le\bar{x}$. Thus, by Remark \ref{rem:32_trailing}, 
\[h(x)\le v_f(x,\bar{x})\le V_{f(\bar{x})}(x),\quad\forall x,\bar{x}\in I\text{ with }x\le \bar{x},\]
\[\left((l,f(\bar{x})]\cap[b(f(\bar{x})),\bar{x}]\right)\equiv\left(\mc{S}_{f(\bar{x})}^{\sf S, L}\cup(l,\bar{x}]\right)=\mc{S}_f^{\sf T, L}(\bar{x}).\]
\item If $f(\bar{x})\ge x_0$, then by Corollary \ref{cor34}, we can use the same argument as above to conclude that $(l,\bar{x}]\equiv\left(\mc{S}_{f(\bar{x})}^{\sf S, L}\cap(l,\bar{x}]\right)=\mc{S}_f^{\sf T, L}(\bar{x})$. 
\end{enumerate}
As a consequence we obtain the following theorem:
\begin{theorem}\label{thm1_trailing}
Under Assumption \ref{assume1}, for $x,\bar{x}\in I$ with $x\le\bar{x}$ and $\bar{x}\ge b_f^\star$, we have 
\[v_f(x,\bar{x})\equiv V_{f(\bar{x})}(x).\]
So the optimal stopping time is $\rho_f\wedge\tau_{X}^+(b(f(\bar{x})))$.
\end{theorem}

In what follows we consider the remaining case $l<x\le \bar{x}<b_f^\star$  and we shall establish the optimality of the stopping rule $\tau_X^+(b_f^\star)\wedge \rho_f$. To this end, we first calculate the associated value of this strategy,   denoted by $u_f(x,\bar{x})$. In particular,  by the strong Markov property of $X$, applying Lemma \ref{fundamentallemma}  we have for any $x\in(f(\bar{x}), \bar{x})$ with $\bar{x}<b_f^\star$, 
\begin{align}
&u_f(x,\bar{x}):=\ex_{x,\bar{x}}(\et^{-r(\rho_f\wedge\tau_X^+(b_f^\star))}h(X_{\rho_f\wedge\tau_X^+(b_f^\star))}))\nn\\
=&h(f(\bar{x}))\ex_x(\et^{-q\tau_X^-(f(\bar{x}))}\ind_{\{\tau_X^-(f(\bar{x}))<\tau_X^+(\bar{x})\}})+u_f(\bar{x},\bar{x})\ex_x(\et^{-q\tau_X^+(\bar{x})}\ind_{\{\tau_X^+(\bar{x})<\tau_X^-(f(\bar{x}))\}})\nn\\
=&\phi_q^-(x)\bigg(\frac{h(f(\bar{x}))}{\phi_q^-(f(\bar{x}))}\frac{\psi_q(\bar{x})-\psi_q(x)}{\psi_q(\bar{x})-\psi_q(f(\bar{x}))}+\frac{u_f(\bar{x},\bar{x})}{\phi_q^-(\bar{x})}\frac{\psi_q(x)-\psi_q(f(\bar{x}))}{\psi_q(\bar{x})-\psi_q(f(\bar{x}))}\bigg),\label{eq14_trailing}
\end{align}
where for $\bar{x}<b_f^\star$, we have
\begin{align}
\frac{u_f(\bar{x},\bar{x})}{\phi_q^-(\bar{x})}=&\frac{h(b_f^\star)}{\phi_q^-(\bar{x})}\ex_{\bar{x},\bar{x}}(\et^{-q\tau_X^+(b_f^\star)}\ind_{\{\tau_X^+(b_f^\star)<\rho_f\}})+\frac{1}{\phi_q^-(\bar{x})}\ex_{\bar{x},\bar{x}}(\et^{-q\rho_f}h(X_{\rho_f})\ind_{\{\rho_f<\tau_X^+(b_f^\star)\}}).\label{eq:excur0}
\end{align}
The two expectations in \eqref{eq:excur0} can be computed  using standard calculation using excursion theory: 
\begin{lemma}\label{trailing:prop_dd}
For any $b>\bar{x}$, we have 
\be
\ex_{\bar{x},\bar{x}}(\et^{-q\rho_f}h(X_{\rho_f})\ind_{\{\rho_f<\tau_X^+(b)\}})
=\phi_q^-(\bar{x})\int_{\bar{x}}^{b}\frac{h(f(v))}{\phi_q^-(f(v))}\frac{\psi_q'(v)}{\psi_q(v)-\psi_q(f(v))}\exp(-\int_{\bar{x}}^v\frac{\psi_q'(u)\,\diff u}{\psi_q(u)-\psi_q(f(u))})\diff v,\nn
\ee
and
\be
\ex_{\bar{x},\bar{x}}(\et^{-q\tau_X^+(b)}\ind_{\{\tau_X^+(b)<\rho_f\}})=\frac{\phi_q^-(\bar{x})}{\phi_q^-(b)}\exp(-\int_{\bar{x}}^b\frac{\psi_q'(u)\diff u}{\psi_q(u)-\psi_q(f(u))}).\nn
\ee
In particular, as $b\to r$ we obtain the value of the plain trailing stop (defined in \eqref{eq:ga})
\be
g_f(\bar{x},\bar{x})=\phi_q^-(\bar{x})\int_{\bar{x}}^{r}\frac{h(f(v))}{\phi_q^-(f(v))}\frac{\psi_q'(v)}{\psi_q(v)-\psi_q(f(v))}\exp(-\int_{\bar{x}}^v\frac{\psi_q'(u)\,\diff u}{\psi_q(u)-\psi_q(f(u))})\diff v,\nn
\ee
and for $f(\bar{x})<x\le \bar{x}$, 
\be
g_f(x,\bar{x})=\phi_q^-(x)\bigg(\frac{(hf(\bar{x}))}{\phi_q^-(f(\bar{x}))}\frac{\psi_q(\bar{x})-\psi_q(x)}{\psi_q(\bar{x})-\psi_q(f(\bar{x}))}+\frac{g_f(\bar{x},\bar{x})}{\phi_q^-(\bar{x})}\frac{\psi_q(x)-\psi_q(f(\bar{x}))}{\psi_q(\bar{x})-\psi_q(f(\bar{x}))}\bigg).\nn
\ee
\end{lemma}

To establish the optimality of $\tau_X^+(b_f^\star)\wedge\rho_f$ when $0<x\le\bar{x}<b_f^\star$, we need to show that the value of the rule $u_f(x,\bar{x})$ dominates the reward function $h(x)$. This claim can be proved  by using \eqref{eq14_trailing} and the optimality of $b_f^\star$ (see Corollary \ref{trailing:cor35}). 

\begin{lemma}\label{lem:36_trailing}
For all $\bar{x}\in(l,b_f^\star)$ and $x\in(f(\bar{x}),\bar{x}]$, we have $u_f(x,\bar{x})>h(\bar{x})$. 
\end{lemma}

Lemma \ref{lem:36_trailing} says that waiting until $\tau_X^+(b_f^\star)\wedge\rho_f$ yields positive ``time value'' $u_f(x,\bar{x})-h(x)>0$ for all $f(\bar{x})<x\le\bar{x}<b_f^\star$, so this region should be part of the optimal continuation region. On the one hand, before hitting $b_f^\star$, this region is obviously the maximum possible continuation region. Furthermore, upon hitting $b_f^\star$ we have $\bar{x}= b_f^\star$, and the case has already been treated in Theorem \ref{thm1_trailing}, which suggest immediate stopping at $\tau_X^+(b_f^\star)$. So we know that the stopping time $\tau_X^+(b_f^\star)\wedge\rho_f$ is optimal for problem \eqref{eq:trailingproblem} if $\bar{x}<b_f^\star$.

\begin{theorem}\label{thm2_trailing}
Under Assumption \ref{assume1}, we have for all $l<x\le\bar{x}<b_f^\star$ that,
\[v_f(x,\bar{x})\equiv u_f(x,\bar{x})=\ex_{x,\bar{x}}(\et^{-q(\tau_X^+(b_f^\star)\wedge\rho_f)})h(X_{\tau_X^+(b_f^\star)\wedge\rho_f}),\]
where $b_f^\star$ is defined in Corollary \ref{trailing:cor35}. Moreover, the mapping $f\mapsto b_f^\star$ is non-increasing over all functions satisfying \eqref{eq:floor}. 
\end{theorem}
\begin{proof}
The only claim that needs a proof is the monotonicity of $f\mapsto b_f^\star$. But that is due to Remark \ref{rem21} and the structure of the optimal stopping region. 
\end{proof}

\begin{corollary}\label{trailing:cor:prem}
The value of the plain trailing stop $g_f(x,\bar{x})$ given in Lemma \ref{trailing:prop_dd} is finite. Moreover, for any $f(\bar{x})<x\le \bar{x}<b_f^\star$, the early liquidation premium given the trailing stop $\rho_f$ is given by 
\begin{align*} p_f(x,\bar{x})=&\frac{\phi_q^-(x)}{\phi_q^-(b_f^\star)}\frac{\psi_q(x)-\psi_q(f(\bar{x}))}{\psi_q(\bar{x})-\psi_q(f(\bar{x}))}\exp(\int_{\bar{x}}^{b_f^\star}\frac{-\psi_q'(u)\diff u}{\psi_q(u)-\psi_q(f(u))})\left(h(b_f^\star)-g_f(b_f^\star, b_f^\star)\right)\nn,\end{align*}
where  $g_f(b_f^\star, b_f^\star)$ is given in Lemma \ref{trailing:prop_dd}.
If $f(\bar{x})<x_0$, $\bar{x}\ge b_f^\star$ and $f(\bar{x})<x<b(f(\bar{x}))$ (see Proposition \ref{trailing:thm33} for the existence of $b(y)$), then the early liquidation premium given the trailing stop $\rho_f$ is 
\begin{align*}
p_f(x,\bar{x})=&\frac{\phi_q^-(x)}{\phi_q^-(b(f(\bar{x})))}\frac{\psi_q(x)-\psi_q(f(\bar{x}))}{\psi(b(f(\bar{x})))-\psi_q(f(\bar{x}))}\left(h(b(f(\bar{x})))-g_f(b(f(\bar{x})),\bar{x})\right).
\end{align*}
Finally, if $f(\bar{x})<x_0$, $\bar{x}\ge b_f^\star$ and $b(f(\bar{x}))\le x\le \bar{x}$, or $f(\bar{x})\ge x_0$ and $f(\bar{x})<x\le\bar{x}$, then the early liquidation premium given the trailing stop $\rho_f$ is 
\begin{align*}
p_f(x,\bar{x})=&h(x)-g_f(x,\bar{x}).
\end{align*}
\end{corollary}

\begin{remark}
If the first inequality in \eqref{eq:H0H} is an equality, then  the optimal threshold $b_f^\star$  
may be at the boundary $r$, in which case, it will be optimal not to liquidate before the trailing stop. That is, $p_f(x,\bar{x})=0$ for all $x,\bar{x}\in I$ such that $x\in(f(\bar{x}),\bar{x}]$.
\end{remark}

\subsection{Optimal Acquisition with a Trailing Stop}
In this section, we solve the optimal stopping problem related to acquisition with a trailing stop, which we recall as follows:
\be
v_f^{(1)}(x)=\sup_{\tau\in\mc{T}}\ex_x(\et^{-q\tau}(v_f(X_\tau,X_\tau)-h_b(X_\tau))\ind_{\{\tau<\infty\}}),\label{eq:compound2}
\ee
where $\mc{T}$ is the set of all stopping times of $X$, and $\sup_{x\in I}(v_f(x,x)-h_b(x))>0$.  

Let us define the optimal acquisition region with a trailing stop as 
\be
\mc{S}_f^{\sf T, A}:=\{x\in I: v_f^{(1)}(x)=v_f(x,x)-h_b(x)\}.\nn
\ee
Following \cite[Proposition 5.10]{DK2003} and \eqref{eq:excur0}, to determine $\mc{S}_f^{\sf T, A}$, it suffices to obtain the smallest concave majorant of 
\be
H^{(1)}(z):=\frac{v_f(x,x)-h_b(x)}{\phi_{q}^-(x)},\quad\text{where }z=\psi_{q}(x)\in\R_+, x\in I.\label{eq:Hfq}\ee
In light of Theorem \ref{thm1_trailing}, we know that for $x\ge b_f^\star$, we have $v_f(x,x)-h_b(x)=h(x)-h_b(x)\le0$, so we must have $\mc{S}_f^{\sf T, A}\subset I\backslash[b_f^\star,r)=(l,b_f^\star)$.  
Therefore, if we denote by 
\be\ol{z}_f^\star:=\sup\mathrm{arg}\max_{z\in\R_+} H^{(1)}(z).\label{eq:zfs}\ee
Then we have $H^{(1)}(\ol{z}_f^\star)>0$ (since $\sup_{x>0}(v_f(x,x)-h_b(x))>0$), and  $\ol{z}_f^\star\in[0,\psi_{q}(b_f^\star))$, and 
the smallest concave majorant of $H^{(1)}(\cdot)$ over $[\ol{z}_f^\star,\infty)$ must be given by the constant function $H^{(1)}(\ol{z}_f^\star)$, so 
we can deduce that $\mc{S}_f^{\sf T,A}\subset(l,\psi_{q}^{-1}(\ol{z}_f^\star)]$. However, no further information about $\mc{S}_f^{\sf T,A}$ is available under general diffusions, mainly due to lack of information about the concavity of $H^{(1)}(\cdot)$.  In fact, as seen in Lemma \ref{trailing:lem:convex} below, even in the special case $h_b(\cdot)\equiv h(\cdot)$, function $H^{(1)}(\cdot)$ over $(0,\psi_q(b_f^\star))$ is the difference between a convex function $H_f(\cdot)$ over $(0,\psi_q(b_f^\star))$ and a function $H(\cdot)$ that is convex over $(0,\psi_q(x_0))$ and is strictly concave over $(\psi_q(x_0),\psi_q(b_f^\star))$, so  we only know that $H^{(1)}(\cdot)$ is convex on $(\psi_q(x_0),\psi_q(b_f^\star))$, but the concavity of this function over $(0,\psi_q(x_0))$ is not available to us. 

\begin{lemma}\label{trailing:lem:convex}
Consider function 
\be
H_f(z):=\frac{v_f(x,x)}{\phi_q^-(x)}, \quad H(z):=\frac{h(x)}{\phi_q^-(x)},\quad \text{where }z=\psi_q(x)\in\mathbb{R}_+.\label{eq:Hdef}\ee
Then $H_f(\cdot)$ is convex on $(0,\psi_q(b_f^\star))$, and $H(\cdot)$ is strictly concave on $(\psi_q(x_0),\infty)$ and is convex on $(0,\psi_q(x_0))$.
\end{lemma}

\begin{remark}
If $X$ follows the  Black-Scholes model with drift $\mu<q$, and volatility $\sigma>0$, then as in \cite{LiLeungWang15}, it  is  never   optimal to acquire the stock given   $h(x)=x-c_s$ and $h_b(x)=x+c_b$, with transaction fees $c_s>0$ and $c_b\ge 0$. To see this, we recall that $v_f(x,x)<V_0(x)=\ind_{\{x< b\}}(\frac{x}{b})^{\beta^+}(b-c_s)+\ind_{\{x\ge b\}}(x-c_s)$, where $\beta^+=\delta+\sqrt{\delta^2+\frac{2q}{\sigma^2}}>1$ with $\delta=\frac{\mu}{\sigma^2}-\half$, and $b=\frac{\beta c_s}{\beta-1}$. The  convexity of $V_0(\cdot)$ implies that $V_0(x)-h(x)<c_s$ for all $x\in\R_+$, so $v_f(x,x)-h(x)<c_s$ for all $x\in\R_+$, for any floor function $f(\cdot)$ that satisfies \eqref{eq:floor}. Thus, we have $v_f(x,x)-h_b(x)=v_f(x,x)-h(x)-(c_b+c_s)<-c_b\le 0$,  so the payoff function for problem \eqref{eq:compound2} to be negative throughout $\R_+$, yielding an empty  optimal stopping region. For some other forms of $h(\cdot)$, one may obtain a non-empty stopping region for problem \eqref{eq:compound2} (see Example \ref{thmF} below). 
\end{remark}
\begin{exa}\label{thmF}
Assuming that $\mu(x)=\mu x, \sigma(x)=\sigma x, h(x)=h_b(x)=x-Kx^{-\epsilon}$ for all $x\in I\equiv\mathbb{R}_+$, where $\mu\in\R$ such that $\mu<q$, and $\sigma,K>0$ and $\epsilon\ge0$ such that $\half\sigma^2\epsilon(\epsilon+1)-\mu\epsilon-q<0$. Let $f(x)=(1-\alpha)x$ for some $\alpha\in(0,1)$.  Then we have $\mc{S}_f^{\sf T,A}=(0,\ul{b}_f^\star]$, where $\ul{b}_f^\star:=\psi_{q}^{-1}(\ol{z}_f^\star)$ with $\ol{z}_f^\star$ given  in \eqref{eq:zfs}, or equivalently, with $\ol{z}_f^\star$  as the unique root to \eqref{eq:trailing:olz} in the Appendix. That is, for all $x\in I$
\[v_f^{(1)}(x)=\ex_x(\et^{-q\tau_X^-(\ul{b}_f^\star)}(v_f(X_{\tau_X^-(\ul{b}_f^\star)},X_{\tau_X^-(\ul{b}_f^\star)})-h_b(X_{\tau_X^-(\ul{b}_f^\star)}))\ind_{\{\tau_X^-(\ul{b}_f^\star)<\infty\}}).\]
\end{exa}

In general, one can analyze the concavity of $H^{(1)}(\cdot)$ (and hence the optimal stopping region) on a case-by-case basis with possibly helps of numerical computation. To demonstrate the idea, let us define 
\be
z_f^\star:=\psi_{q}(b_f^\star),\quad \varphi(z):=\psi_q(f(\psi_q^{-1}(z))),\quad\forall z\in\R_+.\ee
It is clear that $\varphi(\cdot)$ is an increasing function such that $0<\varphi(z)<z$.
From Lemma \ref{trailing:prop_dd} we have for all $z\in(0,z_f^\star)$
\begin{align}
H_f(z)=&\exp(-\int_z^{z_f^\star}\frac{\diff \nu}{\nu-\varphi(\nu)})H(z_f^\star)+\int_z^{z_f^\star}H(\varphi(\nu))\exp(-\int_{z}^{\nu}\frac{\diff w}{w-\varphi(w)})\frac{\diff \nu}{\nu-\varphi(\nu)},\label{eqe29}
\end{align}
where $H_f(\cdot)$ is defined in \eqref{eq:Hdef}.
To obtain the smallest concave majorant of $H^{(1)}(z)=H_f(z)-H(z)-c/\phi_q^-(\psi_q^{-1}(z))$, we need to numerically evaluate $H_f(\cdot)$.
To that end, it will be more convenient to rewrite \eqref{eqe29} into an equivalent first-order linear ODE form:
\be
\begin{dcases}
H_f^{\prime}(z)=\frac{H_f(z)-H(\varphi(z))}{z-\varphi(z)},\quad &\forall z\in(0,z_f^\star),\\
\text{subject to }H_f(z_f^\star)=H(z_f^\star).\label{eqeODE}
\end{dcases}
\ee
Then we can use {\sf Mathematica}'s {\sf NDSolve} command to efficiently   compute the values of $H^{(1)}(\cdot)$ and its derivatives.\footnote{The procedure can be conveniently generalized to allow for distinct discounting rates for the acquisition and liquidation problems.}

%
%

\section{Case Study: Trading  with  a   Trailing Stop under the Exponential OU Model}\label{sec:xou}
In this section, we apply our results in Section \ref{sec:trailing} to an exponential Ornstein-Uhlenbeck (OU) model:
\be
dX_t=X_t\bigg(\lambda(\theta-\log X_t)+\frac{1}{2}\sigma^2\bigg)\diff t+\sigma X_t dW_t,\quad X_0=x\in I\equiv\R_+,\label{eq:XOU}\ee
where $W$ is a standard Brownian motion, $\lambda,\sigma>0$ are positive constants, and $\theta\in\R$ is the long term average for the log-price $\log X$:
\[\diff (\log X_t)=\lambda(\theta-\log X_t)\diff t+\sigma\diff W_t.\]
With reference to \eqref{eq:phi}, it  is well-known (see p.542 of \cite{Borodin2002}) that 
\begin{align*}\phi_q^+(x)=&\et^{\frac{\lambda}{2\sigma^2}(y-\theta)^2}D_{-\frac{q}{\lambda}}(\frac{\sqrt{2\lambda}}{\sigma}(y-\theta)),\\
\phi_q^-(x)=&\et^{\frac{\lambda}{2\sigma^2}(y-\theta)^2}D_{-\frac{q}{\lambda}}(\frac{\sqrt{2\lambda}}{\sigma}(\theta-y)),\end{align*}
where $y=\log x$, and $D_\nu(\cdot)$ is the parabolic cylinder function with parameter $\nu$. 
We are interested in optimal liquidation and acquisition of one unit of an risky asset whose price is modeled by $X$. To that end, we let 
\[h(x)=x-c_0,\quad, h_b(x)=x+c_0,\quad\forall x\in I,\]
where $c_0\ge0$ is a transaction cost to buy or sell. 
Then it follows  that, for any $q>0$
\[(\mc{L}-q)h(x)=\bigg(\lambda(\theta-\log x)+\frac{1}{2}\sigma^2-q\bigg)x+qc_0,\quad\forall x\in I,\]
which is a strictly decreasing function with range equal to $\R$. Moreover, by the asymptotic behavior of $D_{\nu}(\cdot)$ (see e.g. equation (1.8) of \cite{Temme2000}), we know that the reward function $h(\cdot)$ satisfies Assumption \ref{assume1}.  A number of related studies,  such as \cite{Zhang2008,Zervos2013,LeungWang2018}, have also analyzed the optimal buy-low-sell-high strategy under the OU or exponential OU model, with or without a fixed stop-loss exit. Compared to them, we study a different optimal stopping  problem with a random maturity due to the trailing stop. 

\begin{figure}[t!]
\centering
\subfloat[$H_f(z)$ vs. $H(z)$]{{\includegraphics[width=150pt]{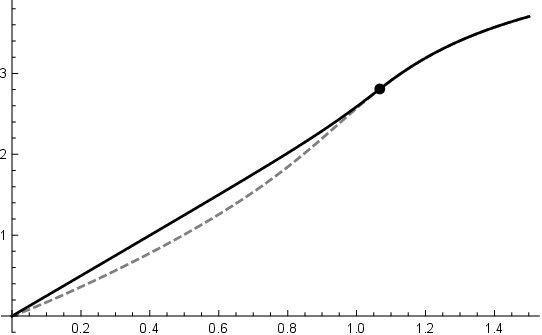} }}%
    \quad
    \subfloat[$v_f(x,x)$ vs. $h(x)$]{{\includegraphics[width=150pt]{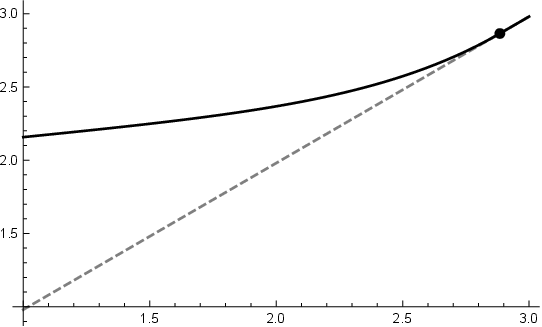} }}\\
    \subfloat[$H^{(1)}(z)$ and its concave majorant]{{\includegraphics[width=150pt]{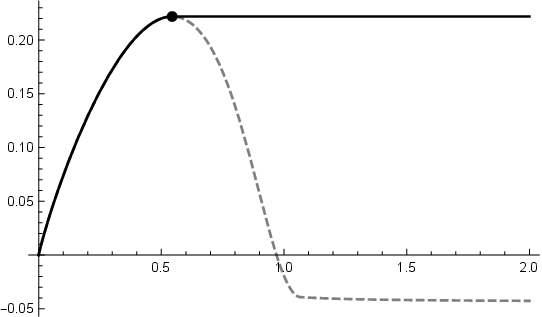} }}%
    \quad
    \subfloat[$v_f^{(1)}(x)$ vs. $v_f(x,x)-h_b(x)$]{{\includegraphics[width=150pt]{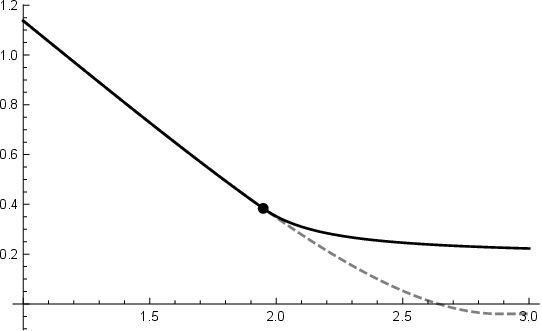} }}
\caption{Numerical results under the exponential OU  model \eqref{eq:XOU}: (a) Plots of function $H(z)$ (dashed gray)  and $H_f(z)$ (solid black). The ``pasting point'' $\psi_q(b_f^\star)=1.0674$ is indicated by  the black dot. (b) Plots of the reward function $h(x)$ (dashed gray)  and the value function $v_f(x,x)$ (solid black). The ``pasting point'' is $b_f^\star=2.8845$ (black dot). (c) Plots of the reward function $H^{(1)}(z)$ (dashed gray)  and its smallest concave majorant  (solid black), along with the  ``pasting point'' $\bar{z}_f^\star=0.5441$ (black dot). (d) Plots of the reward function $v_f(x,x)-h_b(x)$ (dashed gray) and the value function $v_f^{(1)}(x)$ (solid black), and the  ``pasting point'' $\ul{b}_f^\star=1.9488$ (black dot).}\label{fig1_trailing}
\end{figure}

\subsection{Value Function and Optimal Strategy}
Upon purchasing of the asset, we set a percentage drawdown trailing stop, i.e. $f(x)=(1-\alpha)x$, where $\alpha\in(0,1)$ is a constant.

In this study, we select the following parameter values:
\begin{align}
\lambda=0.6,  \theta=1, \sigma=0.2, q=0.05, c_0=0.02, \alpha=0.3.\label{eq:parameter}
\end{align}
This means that we will liquidate the asset whenever its price drops from its running maximum since the acquisition by more than $30\%$.

In Figure \ref{fig1_trailing}(a), we plot the function $H(\cdot)$ defined as in \eqref{eq:Hdef}. We also have plotted 
the function  $H_f(\cdot)$ defined as in \eqref{eq:excur0} (see also \eqref{eqe29}), which is obtained by first solving equation \eqref{eq:cor42} with $f(x)=(1-\alpha)x$ for 
$b_f^\star (=2.8845)$, and then using ODE \eqref{eqe29} to numerically obtain $H_f(\cdot)$.
We notice that, in contrast to the value function for a fixed stop-loss level (Theorem \ref{trailing:thm33}, see also \cite{LiLeung15}), the function $H_f(\cdot)$ is not concave over $(0,\psi_q(b_f^\star))$. This is because, although  $\phi_q^-(x)H_f(\psi_q(x))=v_f(x,x)$ is the value function for the optimal stopping problem \eqref{eq:trailingproblem} when $x=\bar{x}$, it does not yield a martingale of $(X_t,\ol{X}_t)$, which requires using the function $v_f(x,\bar{x})$, not $v_f(x,x)$. 

In Figure \ref{fig1_trailing}(b), we plot the reward function $h(x)$ and the value function $v_f(x,x)$ for the optimal liquidation problem \eqref{eq:trailingproblem} with $x=\bar{x}$.

In Figure \ref{fig1_trailing}(c), we plot the function $H^{(1)}(z)$ defined in \eqref {eq:Hfq} under the current exponential OU model. By checking the function's derivative numerically, we conclude that it is concave to the left of its maximum point. Hence, the smallest concave majorant is given by 
\[\hat{H}_{f,q}^{(1)}(z)=H^{(1)}(z\wedge\ol{z}_f^\star),\quad \forall z\in\R_+.
\]
Therefore, in this case, the optimal acquisition strategy is to purchase the asset once the price is lower than $\ul{b}_f^\star=1.9488$. 

In Figure \ref{fig1_trailing}(d), we plot the function $v_f(x,x)-h_b(x)$ and the value function $v_f^{(1)}(x)$ for the optimal acquisition problem \eqref{eq:trailingproblem2}, and the ``pasting point'' is at $\psi_q^{-1}(\bar{z}_f^\star)=1.9488$. 

In summary, for the exponential OU model \eqref{eq:XOU} with parameters as given in \eqref{eq:parameter}, the optimal trading strategy is to purchase the asset when price is lower than $\psi_q^{-1}(\bar{z}_f^\star)=1.9488$, and setup the $30\%$ trailing stop order as an exit plan, and then wait until either the trailing stop is being activated or the price reaches target $b_f^\star=2.8845$.

\begin{figure}
\centering
\includegraphics[width=3.5in]{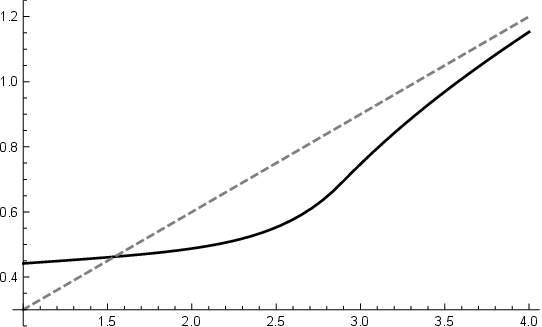}
\caption{Earlier liquidation premium (black) $p_f(x,x)$ and   function $x-f(x)=\alpha x$ (dashed) under  the exponential OU model \eqref{eq:XOU}.}\label{fig:premium_d}
\end{figure}

Lastly, in Figure \ref{fig:premium_d} we plot the early liquidation premium of $\rho_f\wedge\tau_X^+(b_f^\star)$ over the plain trailing stop $\rho_f$ when $x=\bar{x}$. This measure the ``value'' of our result in problem \eqref{eq:trailingproblem}. 
By Corollary \ref{trailing:cor:prem}, we know that, for each $x\in I$, 
\be
p_f(x,x)=\exp(\int_{x}^{b_f^\star\vee x}\frac{-\psi_q'(u)\diff u}{\psi_q(u)-\psi_q(f(u))})\left(h(b_f^\star\vee x)-g_f(b_f^\star\vee x, b_f^\star\vee x)\right).\label{eq:ppp}\ee
To numerically evaluate \eqref{eq:ppp}, we use the fusion of a ``limiting order'' $\tau_X^+(b)$ and the trailing stop $\rho_f$, with $b$ chosen sufficiently large so that 
\begin{eqnarray*}\ex_x(\et^{-q(\tau_X^+(b)\wedge\rho_f)}\ind_{\{\tau_X^+(b)<\rho_f\}})<0.005, \\
 0<h(b)\ex_x(\et^{-q(\tau_X^+(b)\wedge\rho_f)}\ind_{\{\tau_X^+(b)<\rho_f\}})<0.03, \end{eqnarray*}
for all $x$ in the plotting region of Figure \ref{fig:premium_d}. Then $g_f(x,x)$ is approximated by the value of this strategy, which is subsequently solved using an ODE similar as \eqref{eqeODE}.

In Figure \ref{fig:premium_d}, we compare the early liquidation premium $p_f(x,x)$ with the function $x-f(x)=\alpha x$ ($\alpha=0.3$), which is the maximum loss of the trailing stop order if the price $X$ reaches the trailing floor immediately (but without an overshoot).  We notice that, for large $x$, the gain from our strategy over the plain trailing stop approaches 30\% of the price level. Take into account of discounting and transaction costs, this example suggests that setting a trailing stop when the asset price is  high  will almost always incur a 30\% loss at exit. 

\subsection{Sensitivity Analysis and Financial Interpretations}
The following illustrative numerical examples will shed light on  the sensitivity of the optimal acquisition and liquidation thresholds, $\ul{b}_f^\star$ and  $b_f^\star$,  with respect to the trailing stop level $\alpha$, and transaction cost $c_0$. This involve numerical computation of the   thresholds, as well as the critical level where function $(\mc{L}-q)h(x)$ vanishes. 
\begin{figure}
\centering
\subfloat[$(b_f^\star, x_0, \ul{b}_f^\star)$ vs. $\alpha$]{{\includegraphics[width=150pt]{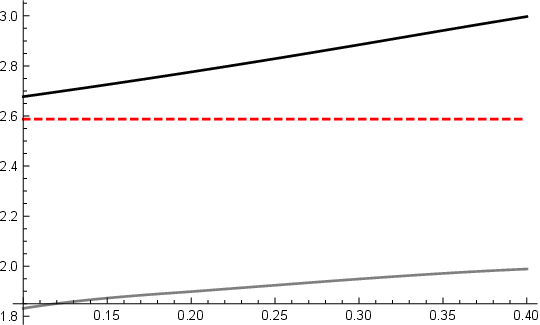} }}%
    \quad
    \subfloat[$(b_f^\star, x_0, \ul{b}_f^\star)$ vs. $\sigma$]{{\includegraphics[width=150pt]{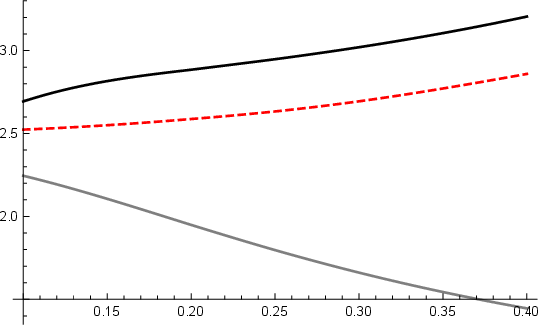} }}\\
    \subfloat[$(b_f^\star, x_0, \ul{b}_f^\star)$ vs. $\lambda$]{{\includegraphics[width=150pt]{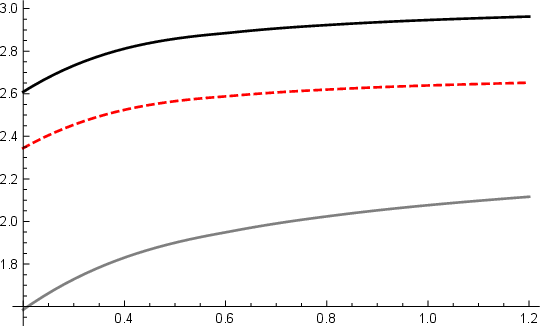} }}%
    \quad
    \subfloat[$(b_f^\star, x_0, \ul{b}_f^\star)$ vs. $c_0$]{{\includegraphics[width=150pt]{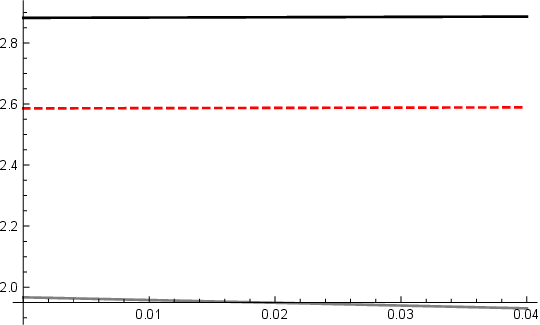} }}
\caption{Sensitivity of thresholds $b_f^\star$ (black), $\ul{b}_f^\star$ (gray), and the root $x_0$ (red dashed) of  $(\mc{L}-q)h(x)=0$,  under  the exponential OU model \eqref{eq:XOU}: (a) Dependence on $\alpha\in[0.1,0.4]$;  (b) Dependence of on $\sigma\in[0.1,0.4]$; (c) Dependence of on $\lambda\in[0.2,1.2]$, and  (d) Dependence of on $c_0\in[0, 0.04]$.  In all figures, other parameters  are set as in \eqref{eq:parameter}.}\label{fig:sensitivity}
\end{figure}
In Figure \ref{fig:sensitivity}(a), we plot $(b_f^\star,x_0,\ul{b}_f^\star)$ as a function of the trailing stop   level $\alpha$, with  $x_0$ (the dashed line)   defined in Assumption \ref{assume1}. The optimal liquidation level $b_f^\star$ is increasing in $\alpha$, confirming our result in Theorem \ref{thm2_trailing}. Moreover, the optimal acquisition level $\ul{b}_f^\star$ is also increasing in $\alpha$. Recalling that a higher $\alpha$ means a lower trailing stop trigger, this means that a larger downside protection   induces the investor to enter the market earlier. As seen   in Figure \ref{fig:sensitivity}(a), the investor with a higher $\alpha$ will acquire the asset at a price level closer to the critical level $x_0$.  Our numerical results also suggest  that, for small $\alpha$, it may not be optimal to initiate the position at all, because  the gain to be realized at the sell order at $b_f^\star$ or at the trailing stop   will be  too low   compared to  the transaction cost $c_0$. In such cases, we observe that  $\sup_{x\in\R}(v_f(x,x)-h(x))<c_0=0.02$. 

In Figure \ref{fig:sensitivity}(b), we plot $(b_f^\star,x_0,\ul{b}_f^\star)$  as a function of the asset's volatility parameter $\sigma$. We see that, as $\sigma$ increases, the optimal liquidation level increases, thanks to stronger force from the Brownian motion. However, the acquisition price level is lower for higher  $\sigma$, which means that the investor is willing to establish a position at a lower price. However, higher volatility will increase the likelihood for the asset price to reach low levels earlier, so the actual entry time by the investor may be earlier or later. The decreasing pattern of $\ul{b}_f^\star$ with respect to $\sigma$ suggests that the investor   voluntarily lowers the take-profit level to mitigate the risk of realizing a reduced profit or  a loss at the  trailing stop in a more volatile market.

 Figure  \ref{fig:sensitivity}(c) illustrates the effect of  the asset's rate of mean reversion $\lambda$. A higher  $\lambda$ means that    the log-price will move around its long-term mean $\theta$ faster. As a response, the investor enters the market earlier at a higher entry level and exit at a lower level, resulting in a quick roundtrip, as reflected in the plot by the increasing trends of  $b_f^\star$ and $\ul{b}_f^\star$ with respect to   $\lambda$. Moreover, their distance is shrinking as $\lambda$ continues to increase. Intuitively, since  the asset price tends to rapidly revert  to the mean, it does not make sense to select entry and exit price levels that are far apart and away from the mean as the chance of execution is too low. 
 
%

The effect of transaction cost $c_0$ is shown in   Figure \ref{fig:sensitivity}(d), where we plot  $(b_f^\star,x_0,\ul{b}_f^\star)$  as a function of  $c_0$. The optimal liquidation   level $b_f^\star$ increases slightly with respect to $c_0$ while the optimal acquisition level $\ul{b}_f^\star$ decreases in $c_0$. To interpret,    higher transaction costs discourage both acquisition and liquidation, though the effect is not significant. Nevertheless,  as pointed out  in our analysis above, while there is always a finite optimal liquidation price $b_f^\star$ given  any transaction cost, a high transaction cost may make the trade unprofitable and thus exclude market entry.

\appendix
\section{Proofs}
\begin{proof}[Proof of Lemma \ref{lem:threshold_opt}]
Following \cite{DK2003}, let us define 
For any $b\in I$, let us define 
\be
H(z):=\frac{h(x)}{\phi_q^-(x)},\text{ where }z=\psi_q(x)\in \mathbb{R}_+.\label{eq:def_Hz}
\ee
By \cite[Proposition 5.11]{DK2003}, we know that the value function 
\[V(x):=\sup_{\tau\in\mc{T}}\ex_x(\et^{-r\tau}h(X_\tau)\ind_{\{\tau<\infty\}}),\]
is given by $\phi_q^-(x)\hat{H}(\psi_q(x))$, where $\hat{H}(\cdot)$ is the smallest nonnegative concave majorant of $H(\cdot)$ on $\mathbb{R}_+$. On the other hand, by \cite[Section 6]{DK2003}, we have
\[H''(z)=\frac{2}{\sigma^2(x)\phi_q^-(x)(\psi_q'(x))^2}\left((\mc{L}-q)h(x)\right), \quad\text{for }z=\psi_q(x).\]
So Assumption \ref{assume1} implies that $H(\cdot)$ is convex on $(0,\psi_q(x_0))$, and concave on $(\psi_q(x_0),\infty)$. 
We now examine the behavior of $H(\cdot)$ near 0 and $\infty$. From \eqref{eq:def_Hz} we know that, 
\begin{enumerate}
\item if $h(l+)\ge0$, then $h(l+)$ is finite, and  $H(0+)=\lim_{x\downarrow l}\frac{h(x)}{\phi_q^-(x)}=0$;
\item if $h(l+)<0$, then $H(z)<0$ for sufficiently small $z>0$.
\end{enumerate} 
Moreover, from
\[F(z):=\frac{H(z)}{z}=\frac{h(x)}{\phi_q^+(x)},\text{ where }z=\psi_q(x)\]
we know $H(z)>0$ for sufficiently large $z>0$. Here, function $F(\cdot)$ is twice continuously differentiable on $\R_+$, and by Assumption \ref{assume1}  we know that $\sup_{z\ge \psi_q(x_0)}F(z)=F(z_*)$ for some $z_*\in[\psi_q(x_0),\infty)$. Obviously $F(z_*)>0$, which implies that $H(z)=\frac{h(x)}{\phi_q^-(x)}>0$ for all $z>z_*$ since $h(\cdot)$ is monotone. Furthermore, $z_*$ must  satisfy the first order condition 
\be
\frac{1}{z_*}(H'(z_*)-F(z_*))=0.\label{eq:first_order_z}
\ee
 Now define function 
\[\tilde{H}(z)=F(z_*)z\ind_{\{z<z_*\}}+H(z)\ind_{\{z\ge z_*\}},\]
which is clearly continuously differentiable and concave on $\mathbb{R}_+$, thanks to \eqref{eq:first_order_z}. Function $\tilde{H}(\cdot)$ is also positive on $\mathbb{R}_+$, which is evident from the construction. Hence we conclude that $\tilde{H}(\cdot)$ is the smallest concave majorant of $H(\cdot)$. So the optimal stopping region is given by 
\[\psi_q^{-1}(\{z\in\R_+: \tilde{H}(z)=H(z)\})=(\psi_q^{-1}(z_*),r).\]
Therefore, $x^\star=\psi_q^{-1}(z_*)$ is the optimal stopping threshold. 
\end{proof}

\begin{proof}[Proof of Proposition \ref{trailing:thm33}]
The proof is similar as that for Lemma \ref{lem:threshold_opt}. In the spirit of \cite{DK2003}, we derive the optimal value function and the stopping region by constructing  the smallest concave majorant of $H(z)$ on $[\psi_q(y),\infty)$. By the convexity of $H(\cdot)$, we know this concave majorant is given by 
\be\hat{H}_y(z)
=\begin{dcases}
H(\psi_q(y))\frac{z(y)-z}{z(y)-\psi_q(y)}+H(z(y))\frac{z-\psi_q(y)}{z(y)-\psi_q(y)}, &\forall z\in(\psi_q(y),z(y)),\\
H(z), &\forall z\not\in(\psi_q(y),z(y)),
\end{dcases}\label{eq:Hhat}\ee
where $z(y)$ is defined as
\be
\label{eq:by1}z(y):=\inf\mathop{\arg\max}_{z>\psi_q(x_0)}\frac{H(z)-H(\psi_q(y))}{z-\psi_q(y)}.\ee
Thus, the optimal stopping region is given by 
\[\mc{S}_y^{\sf S,L}=\psi_q^{-1}(\R_+\backslash(\psi_q(y),z(y)))=(l,y]\cup[\psi_q^{-1}(z(y)),r).\]
Therefore, the optimal stopping barrier is given by 
 $b(y):=\psi_q^{-1}(z(y))$. 
 
From Remark \ref{rem:mono} we know that, for $l\le y_1<y_2<x_0$, the equalities hold:
\[\left((l,y_1]\cup[b(y_1),r)\right)\equiv \mc{S}_{y_1}^{\sf S, L}\subset\mc{S}_{y_2}^{\sf S, L}\equiv \left((l,y_2]\cup[b(y_2),r)\right).\]
Thus necessarily, $b(y_2)\le b(y_1)\le b(l)=x_*<r$. Because $z(y)$ is an interior maximizer in the objective function in \eqref{eq:by1}, it must satisfy the first order condition:
\be
\frac{1}{z(y)-\psi_q(y)}\bigg(H'(z(y))-\frac{H(z(y))-H(\psi_q(y))}{z(y)-\psi_q(y)}\bigg)=0.\label{eq:first_order_zy}
\ee
This gives \eqref{trailing:eq:8.20}.
%

As $y\uparrow x_0$, $b(y)$ converges to some limit in $[x_0,r)$. Suppose  that  $b(x_0-)\equiv\underline{b}>x_0$, then   the concavity of $H(\cdot)$ over $(\psi_q(x_0),\infty)$ implies that 
\be
H'(\psi_q(\underline{b}))\le \frac{H(\psi_q(\underline{b}))-H(\psi_q(x_0))}{\psi_q(\underline{b})-\psi_q(x_0)}.\nn\ee
However, taking limit in \eqref{eq:first_order_zy} as $y\uparrow x_0$, we know that the above  inequality is in fact an equality. 
This, together with the concavity of $H(\cdot)$ implies that $H(\cdot)$ is in fact a straight line over $[\psi_q(x_0), \psi_q(\underline{b})]$, but then (by the definition of $z(y)$, again) we 
must  have $b(x_0-)=x_0$ instead.

We use implicit differentiation to prove   $b(y)$ is strictly decreasing and differentiable on $(l,x_0)$. To that end, we denote $z=z(y)$ and $u=\psi_q(y)$, then the first order equation in \eqref{eq:first_order_zy} reads as
\[f(z,u)=0, \text{ where }f(z,w)=H'(z)-\frac{H(z)-H(u)}{z-u}.\]
By the definition of $z\equiv z(y)$ we have 
\begin{align*}\frac{\partial f}{\partial u}&=H'(u)-\frac{H(z)-H(u)}{(z-u)}<0,\\
 \frac{\partial f}{\partial z}&=H''(z)-\frac{1}{z-u}f(z,u)=H''(z)<0.\end{align*}
Thus, we know that $z(y)$ is strictly decreasing and differentiable in $\psi_q(y)$. In order words, $z(y)$ is differentiable in $y$ and $z'(y)<0$ for any $y\in(l,x_0)$.  
 \end{proof}

\begin{proof}[Proof of Corollary \ref{trailing:cor35}]
From Theorem \ref{trailing:thm33} we know that $\bar{x}\mapsto b(f(\bar{x}))$ is strictly decreasing and continuous over $(f^{-1}(l),f^{-1}(x_0))$, and the mapping $\bar{x}:\mapsto\bar{x}$ is strictly increasing over the same domain. Therefore,  the difference $D(\bar{x}):=b(f(\bar{x}))-\bar{x}$ is strictly decreasing, and $D(\bar{x})\ge D(x_0)> 0$ for all $\bar{x}\in(f^{-1}(l),x_0]$, and by Proposition \ref{trailing:thm33},
\[ \lim_{\bar{x}\uparrow f^{-1}(x_0)}D(\bar{x})=x_0-f^{-1}(x_0)<0.\]
As a consequence, we can define $b_f^\star:=\inf\{\bar{x}<f^{-1}(x_0): D(\bar{x})\le 0\}$, and $b_f^\star\in(x_0,f^{-1}(x_0))$,  so $f(b_f^\star)\le x_0$.

Now for all $\bar{x}<b_f^\star$, by the construction of $b_f^\star$ we have $b(f(\bar{x}))>\bar{x}$, by definition of $z(f(\bar{x}))\equiv \psi_q(b(f(\bar{x})))$ in the proof of Proposition \ref{trailing:thm33} we know that $z(f(\bar{x}))>\psi_q(\bar{x})$. Because the line segment $l_0$ connecting $(\psi_q(f(\bar{x})), H(\psi_q(f(\bar{x}))))$ and $(z(f(\bar{x})), H(\psi_q(f(\bar{x}))))$ gives part of the concave majorant of  $H(\cdot)$,   
we know that the line segment $l_1$ connecting $(\psi_q(f(\bar{x})), H(\psi_q(f(\bar{x}))))$ and $(\psi_q(\bar{x}), H(\psi_q(\bar{x})))$, which is below line segment $l_0$,  must go below the graph of $H(\cdot)$ at $\psi_q(\bar{x})$. This implies that the derivative of $H(\cdot)$ at $\psi_q(\bar{x})$ must be strictly greater than that of line segment $l_1$. That is, 
\[H'(\psi_q(\bar{x}))>\frac{H(\psi_q(\bar{x}))-H(\psi_q(f(\bar{x})))}{\psi_q(\bar{x})-\psi_q(f(\bar{x}))}\quad\Leftrightarrow\quad\Gamma(\bar{x})>0.\]
On the other hand, for all $f^{-1}(x_0)>\bar{x}>b_f^\star$, we have $b(f(\bar{x}))<\bar{x}$. Using similar argument as above, we know that $z(f(\bar{x}))=\psi_q(b(f(\bar{x})))<\psi_q(\bar{x})$. Since the line segment $l_1$ connecting  $(\psi_q(f(\bar{x})), H(\psi_q(f(\bar{x}))))$ and $(\psi_q(\bar{x}), H(\psi_q(\bar{x})))$ is a line segment connecting two points on the graph of a concave function $\hat{H}(\cdot)$, which is the smallest concave majorant of $H(\cdot)$ over $[\psi_q(f(\bar{x})),\infty)$, we know that 
\[\hat{H}'(\psi_q(\bar{x}))=H'(\psi_q(\bar{x}))<\frac{H(\psi_q(\bar{x}))-H(\psi_q(f(\bar{x})))}{\psi_q(\bar{x})-\psi_q(f(\bar{x}))}\quad\Leftrightarrow\quad\Gamma(\bar{x})<0.\]
Expressing $H(\cdot)$ and its derivative with $h(\cdot), \phi_q^-(\cdot), \psi_q(\cdot)$ and their derivatives yields \eqref{eq:cor42} and completes the proof.
\end{proof}

 \begin{proof}[Proof of Lemma \ref{trailing:prop_dd}]
Let us denote by $\e_q$ an exponential random variable with mean $1/q$, which is independent of $X$. Then we notice that 
\begin{align*}
\ex_{\bar{x},\bar{x}}(\et^{-q\rho_f}h(X_{\rho_f})\ind_{\{\tau_X^+(b)<\rho_f\}})=&\ex_{\bar{x},\bar{x}}(h(X_{\rho_f})\ind_{\{\rho_f<\tau_X^+(b)\wedge\e_q\}}),\\
\ex_{\bar{x},\bar{x}}(\et^{-q\tau_X^+(b)}\ind_{\{\tau_X^+(b)<\rho_f\}})=&\pr_{\bar{x},\bar{x}}(\tau_X^+(b)<\rho_f\wedge\e_q).
\end{align*}
To calculate the right-hand sides of the above, we consider an excursion of $X$ below $u$ (notice that $\tau_X^+(u-)=\inf\{t>0: X_t\ge u\}$ is the first hitting time of $X$ to $u$):
\[\es_u=\{\es_u(s):=X_{\tau_X^+(u-)}-X_{\tau_{X}^+(u-)+s}\}_{0<s\le \tau_X^+(u)-\tau_X^+(u-)},\]
which is defined for all $u\ge X_0=\ol{X}_0=\bar{x}$ such that its lifetime $\zeta(\es_u):=\tau_X^+(u)-\tau_X^+(u-)>0$. When $\zeta(\es_u)=0$ we set $\es_u=\partial$, an isolated point. Then the process $\{(u,\es_u)\}_{u\ge\bar{x}}$ is a Poisson point process with jump measure $\diff u\times\diff n_u$, where $n_u$ is the excursion measure for $\es_u$. 
Define $T_f(\es_u):=\inf\{0<s<\zeta(\es_u): \es_{u}(s)>u-f(u)\}$.  It is known from \cite{Salminen2007a} and Lemma \ref{fundamentallemma} that, 
\begin{align}
n_u(\e_q<\zeta(\es_u)\wedge T_f(\es_u))=&\lim_{x\uparrow u}\frac{1}{u-x}\left(1-\ex_{x}(\et^{-q\tau_X^+(u)}\ind_{\{\tau_X^+(u)<\tau_X^-(f(u))\}})\right)-\lim_{x\uparrow u}\frac{\ex_{x}(\et^{-q\tau_X^-(f(u))}\ind_{\{\tau_X^-(f(u))<\tau_X^+(u)\}})}{u-x}\nn\\
=&\frac{\phi_q^{-,\prime}(u)}{\phi_q^-(u)}+\bigg(1-\frac{\phi_q^-(u)}{\phi_q^-(f(u))}\bigg)\frac{\psi_q'(u)}{\psi_q(u)-\psi_q(f(u))},\nn\end{align}
\begin{align}
n_u(T_f(\es_u)<\zeta(\es_u)\wedge\e_q)
=&\lim_{x\uparrow u}\frac{\ex_{x}(\et^{-q\tau_X^-(f(u))}\ind_{\{\tau_X^-(f(u))<\tau_X^+(u)\}})}{u-x}=\frac{\phi_q^-(u)}{\phi_q^-(f(u))}\frac{\psi_q'(u)}{\psi_q(u)-\psi_q(f(u))}.\nn
\end{align}
Hence, 
\[n_u(\e_q<\zeta(\es_u)\wedge T_f(\es_u)\text{ or }T_f(\es_u)<\zeta(\es_u)\wedge\e_q)=\frac{\phi_q^{-,\prime}(u)}{\phi_q^-(u)}-\frac{\psi_q'(u)}{\psi_q(u)-\psi_q(f(u))}.
\]
Let $A$ be the space of all excursions $\es_u$ such that $T_f(\es_u)<\zeta(\es_u)\wedge\e_q$, and $B$ be the space of all excursions $\es_u$ such that $\e_q<\zeta(\es_u)\wedge T_f(\es_u)$. We have that $A\cap B=\emptyset$. Consider a Poisson process (with time indexed by the running maximum $\ol{X}$) that jumps whenever the current excursion $\es_{\ol{X}}\in A\cup B$,  then from the above calculation, we know that this Poisson process has jump intensity $n_u(\e_q<\zeta(\es_u)\wedge T_f(\es_u)\text{ or }T_f(\es_u)<\zeta(\es_u)\wedge\e_q)$. So 
$\pr_{\bar{x},\bar{x}}(\tau_X^+(b)<\rho_f\wedge\e_q)$ is the same as the probability that this Poisson process has no jump over $[\bar{x},b)$, which is given by
\begin{align}
\exp(-\int_{\bar{x}}^bn_u(\e_q<\zeta(\es_u)\wedge T_f(\es_u)\text{ or }T_f(\es_u)<\zeta(\es_u)\wedge\e_q)\diff u)=&\frac{\phi_q^-(\bar{x})}{\phi_q^-(b)}\exp(-\int_{\bar{x}}^b\frac{\psi_q'(u)\diff u}{\psi_q(u)-\psi_q(f(u))}).\nn
\end{align}

Moreover, for any $v\in[\bar{x},b)$, the probability that the Poisson process will have the first jump at ``time'' $\diff v$ as a result of $\es_v\in A$, is given by 
\begin{align*}
&\exp(-\int_{\bar{x}}^vn_u(\e_q<\zeta(\es_u)\wedge T_f(\es_u)\text{ or }T_f(\es_u)<\zeta(\es_u)\wedge\e_q)\diff u)\cdot n_v(T_f(\es_v)<\zeta(\es_v)\wedge\e_q)\diff v\nn\\
=&\frac{\phi_q^-(\bar{x})}{\phi_q^-(v)}\exp(-\int_{\bar{x}}^v\frac{\psi_q'(u)\diff u}{\psi_q(u)-\psi_q(f(u))})\times\frac{\phi_q^-(v)}{\phi_q^-(f(v))}\frac{\psi_q'(v)}{\psi_q(v)-\psi_q(f(v))}\diff v\nn\\
=&\frac{\phi_q^-(\bar{x})}{\phi_q^-(f(v))}\frac{\psi_q'(v)}{\psi_q(v)-\psi_q(f(v))}\exp(-\int_{\bar{x}}^v\frac{\psi_q'(u)\diff u}{\psi_q(u)-\psi_q(f(u))})\diff v\nn,
\end{align*}
which is the same as $\pr_{\bar{x},\bar{x}}(\ol{X}_{\rho_f}\in\diff v, \rho_f<\tau_X^+(b)\wedge\e_q)$. The proof is complete by integrating in $v$ over $[\bar{x},b)$.
\end{proof}

\begin{proof}[Proof of Lemma \ref{lem:36_trailing}]
Let us define for any $b\ge \bar{x}$
\begin{multline}
\bar{H}(\psi_q(\bar{x}), b):=H(\psi_q(b))\exp(-\int_{\bar{x}}^{b}\frac{\psi_q'(u)\,\diff u}{\psi_q(u)-\psi_q(f(u))})\nn\\
+\int_{\bar{x}}^{b}\frac{\psi_q'(v)\,H(\psi_q(f(v)))}{w(v)-w(f(v))}\exp(-\int_{\bar{x}}^{v}\frac{\psi_q'(u)}{\psi_q(u)-\psi_q(f(u))}\diff u)\diff v.
\end{multline}
It is clear that $\bar{H}(\psi_q(\bar{x}),\bar{x})=H(\psi_q(\bar{x}))=\frac{h(\bar{x})}{\phi_q^-(\bar{x})}$, and for $b>\bar{x}$ we have the right derivative of $H_f(\psi_q(\bar{x}),b)$ in $b$:
\begin{align}
&\frac{\partial}{\partial b}\bar{H}(\psi_q(\bar{x}), b)\nn\\
=&\psi_q'(b)\exp(-\int_{\bar{x}}^{b}\frac{\psi_q'(u)\diff u}{\psi_q(u)-\psi_q(f(u))})\bigg(H_+'(\psi_q(b))-\frac{H(\psi_q(b))-H(\psi_q(f(b)))}{\psi_q(b)-\psi_q(f(b))}\bigg).\nn
\end{align}
It follows that the sign of $\frac{\partial}{\partial b}\bar{H}(\psi_q(\bar{x}), b)$ depends on that of
\be
\Gamma(b)=H'(\psi_q(b))-\frac{H(\psi_q(b))-H(\psi_q(f(b)))}{\psi_q(b)-\psi_q(f(b))}.\nn\ee
But the latter is known to be positive for all $b<b_f^\star$, thanks to Corollary \ref{trailing:cor35}. Because $H'(\psi_q(\cdot))$ is continuous, so is $\Gamma(\cdot)$. So we know that 
\begin{align*}\frac{u_f(x,\bar{x})}{\phi_q^-(x)}=&\bar{H}(\psi_q(\bar{x}),b_f^\star)=H(\psi_q(\bar{x}))+\int_{\bar{x}}^{b_f^\star}\frac{\partial}{\partial u}\bar{H}(\psi_q(\bar{x}), u)\diff u>H(\psi_q(\bar{x}))=\frac{h(x)}{\phi_q^-(x)}, \forall \bar{x}<b_f^\star.\end{align*}
This completes the proof.
\end{proof}

\begin{proof}[Proof of Corollary \ref{trailing:cor:prem}]
If $f(\bar{x})<x\le\bar{x}<b_f^\star$, then by the strong Markov property of $X$, we have
\begin{align}
p_f(x,\bar{x})=&\ex_{x,\bar{x}}([\et^{-q\tau_X^+(b_f^\star)}h(X_{\tau_X^+(b_f^\star)})-\et^{-q\rho_f}h(X_{\rho_f})]\ind_{\{\tau_X^+(b_f^\star)<\rho_f<\infty\}})\nn\\
=&\ex_{x,\bar{x}}(\et^{-q\tau_X^+(b_f^\star)}\ind_{\{\tau_X^+(b_f^\star)<\rho_f\}})\, \left(h(b_f^\star)-\ex_{b_f^\star,b_f^\star}(\et^{-q\rho_f}h(X_{\rho_f})\ind_{\{\rho_f<\infty\}})\right),\nn
\end{align}
where  $\ex_{b_f^\star,b_f^\star}(\et^{-q\rho_f}h(X_{\rho_f})\ind_{\{\rho_f<\infty\}})=g_f(b_f^\star,b_f^\star)$ is given in Lemma \ref{trailing:prop_dd}, which is finite since we know that it is dominated from above by $v_f(b_f^\star,b_f^\star)=h(b_f^\star)$. On the other hand, by the analysis in \eqref{eq14_trailing} and the results in Lemma \ref{trailing:prop_dd}, we have
\begin{align*}\ex_{x,\bar{x}}(\et^{-q\tau_X^+(b_f^\star)}\ind_{\{\tau_X^+(b_f^\star)<\rho_f\}})=&\frac{\phi_q^-(x)}{\phi_q^-(b_f^\star)}\frac{\psi_q(x)-\psi_q(f(\bar{x}))}{\psi_q(\bar{x})-\psi_q(f(\bar{x}))}\exp(-\int_{\bar{x}}^{b_f^\star}\frac{\psi_q'(u)\diff u}{\psi_q(u)-\psi_q(f(u))})\nn.\end{align*}
We obtain the claimed formula by combining the above results.

If $f(\bar{x})<x_0$ and $\bar{x}\ge b_f^\star$, then from Theorem \ref{trailing:thm33} and Theorem \ref{thm1_trailing} we know that $b(f(\bar{x}))\le \bar{x}$, and for all $f(\bar{x})<x<b(f(\bar{x}))$,
\begin{align*}
p_f(x,\bar{x})=&\ex_x(\et^{-q\tau_X^+(b(f(\bar{x})))}\ind_{\{\tau_X^+(b(f(\bar{x})))<\tau_X^-(f(\bar{x}))\}})\left(h(b(f(\bar{x})))-\ex_{b(f(\bar{x}),\bar{x}}(\et^{-q\rho_f}h(X_{\rho_f})\ind_{\{\rho_f<\infty\}})\right).
\end{align*}
By using Lemma \ref{fundamentallemma}  we obtain that
\begin{align*}\ex_x(\et^{-q\tau_X^+(b(f(\bar{x})))}\ind_{\{\tau_X^+(b(f(\bar{x})))<\tau_X^-(f(\bar{x}))\}})=&\frac{\phi_q^-(x)}{\phi_q^-(b(f(\bar{x})))}\frac{\psi_q(x)-\psi_q(f(\bar{x}))}{\psi(b(f(\bar{x})))-\psi_q(f(\bar{x}))}.\end{align*}

The claim in this case follows from Lemma \ref{trailing:prop_dd}. 

In the last case that $f(\bar{x})<x_0$, $\bar{x}\ge b_f^\star$ and $b(f(\bar{x}))\le x\le \bar{x}$, or $f(\bar{x})\ge x_0$ and $f(\bar{x})<x\le\bar{x}$, from Theorem \ref{trailing:thm33} and Theorem \ref{thm1_trailing} we know that the optimal stopping rule for problem \eqref{eq:trailingproblem} is 0, so we have
\[p_f(x,\bar{x})=h(x)-\ex_{x,\bar{x}}(\et^{-q\rho_f}h(X_{\rho_f})\ind_{\{\rho_f<\infty\}}).\]
The completes the proof.
\end{proof}

\begin{proof}[Proof of Lemma \ref{trailing:lem:convex}]
The convexity of $H(\cdot)$ has already been proved in the proof of Lemma \ref{lem:threshold_opt}, so we only need to prove that for $H_f(\cdot)$. To that end, we recall \eqref{eqeODE} that 
\[H_f^{\prime}(z)=\frac{H_f(z)-H(\varphi(z))}{z-\varphi(z)},\quad \forall z\in(0,z_f^\star),\]
from which we obtain that, for $z\in(0,z_f^\star)$,
\begin{align}
\diff H_f^{\prime}(z)&=\frac{\frac{H_f(z)-H(\varphi(z))}{z-\varphi(z)}\diff z-H'(\varphi(z))\diff\varphi(z)}{z-\varphi(z)}-\frac{H_f(z)-H(\varphi(z))}{(z-\varphi(z))^2}(\diff z-\diff\varphi(z))\nn\\
&=\bigg(\frac{H_f(z)-H(\varphi(z))}{z-\varphi(z)}-H'(\varphi(z))\bigg)\frac{\diff\varphi(z)}{z-\varphi(z)}\nn\\
&\ge \bigg(\frac{H(z)-H(\varphi(z))}{z-\varphi(z)}-H'(\varphi(z))\bigg)\frac{\diff\varphi(z)}{z-\varphi(z)}.\label{trailing:eq:38}
\end{align}
We prove that the embraced expression in \eqref{trailing:eq:38} is positive, which implies that $H_f'(\cdot)$ is increasing so $H_f(\cdot)$ is convex. 

To prove the claim, we notice that for $z\in(0,z_f^\star)$, we have $\varphi(z)<\varphi(z_f^\star)=\psi_q(f(\psi_q^{-1}(z_f^\star)))=\psi_q(f(b_f^\star))<\psi_q(x_0)$, thanks to Corollary \ref{trailing:cor35}. We now prove that the line segment connecting $(\varphi(z),H(\varphi(z)))$ and $(z,H(z))$ stays above the graph of $H(\cdot)$. Suppose not, then by the convexity of $H(\cdot)$ this can happen only if the line segment crosses the graph of $H(\cdot)$ twice, and $z>\psi_q(b(\psi_q^{-1}(\varphi(z))))$, the latter of which is the point where the tangent line of $H(\cdot)$ that crosses $(\varphi(z),H(\varphi(z)))$ touches the graph of $H(\cdot)$. In other words, 
\be\psi_q^{-1}(z)>b(\psi_q^{-1}(\varphi(z)).\label{trailing:eq:39}\ee
On the other hand, by the monotonicity of $b(y)$ (see Proposition \ref{trailing:thm33}) we know that 
\be
b(\psi_q^{-1}(\varphi(z))>b(\psi_q^{-1}(\varphi(z_f^\star))=b_f^\star,\label{trailing:eq:40}
\ee
where we used the definition of $b_f^\star$ in Corollary \ref{trailing:cor35}. However, \eqref{trailing:eq:39} is contradictory to \eqref{trailing:eq:40}. Thus, the the line segment connecting $(\varphi(z),H(\varphi(z)))$ and $(z,H(z))$ stays above the graph of $H(\cdot)$. Given that $H(\cdot)$ is convex at $\varphi(z)$, we know that the slope of this line segment, $\frac{H(z)-H(\varphi(z))}{z-\varphi(z)}$, is larger than $H'(\varphi(z))$.
\end{proof}

\begin{lemma}\label{lemmabeta}Define the constant $\beta^\pm :=  -\delta\pm\gamma,$ where\[\delta=\frac{\mu}{\sigma^2}-\half,\quad \gamma=\sqrt{\delta^2+\frac{2q}{\sigma^2}}.\]
Then, we have $\beta^+>1$ and 
\be
\frac{-\epsilon-\beta^-}{2\gamma}, \quad \frac{1-\beta^-}{2\gamma}\in(0,1).\label{trailing:eq:42}
\ee
\end{lemma}
\begin{proof}First, since $g(1)=\mu-q<g(\beta^+)=0$ where $g(\beta)=\half\sigma^2\beta(\beta-1)+\mu\beta-q$, we conclude that $1<\beta^+$. It follows from $\delta<\gamma$ that  $-\beta^-=\delta+\gamma<2\gamma$, so $-\frac{\beta^-}{2\gamma}<1$. From $g(-\epsilon)<g(\beta^-)=0$ where $g(\beta)=\half\sigma^2\beta(\beta-1)+\mu\beta-q$, we know that $-\epsilon>\beta^-$. Moreover, $1-\beta^--2\gamma=1+\delta-\gamma=1+\delta-\sqrt{\delta^2+\frac{2q}{\sigma^2}}<\delta+1-\sqrt{\delta^2+\frac{2\mu}{\sigma^2}}=\delta+1-\sqrt{\delta^2+2\delta+1}\le0$, so $\frac{1-\beta^-}{2\gamma}<1$. 
\end{proof}
\begin{proof}[Proof of Example \ref{thmF}]
First of all, we verify that $h(\cdot)$ satisfies  Assumption \ref{assume1}. To that end, we calculate
\[(\mc{L}-q)h(x)=(\mu-q)x-[\half\sigma^2\epsilon(1+\epsilon)-\mu\epsilon-q] Kx^{-\epsilon},\]
from which we know that \eqref{eq:trailing:assume1} holds. 
From \cite{Borodin2002} we know that 
\be\phi_q^\pm(x)=x^{\beta^\pm}, \psi_q(x)=x^{\beta^+-\beta^-}=x^{2\gamma},\label{trailing:eq:41}\ee
where $\beta^\pm$ is defined in Lemma \ref{lemmabeta}.  Condition \eqref{eq:H0H} holds since $\beta^+>1$, and thus,   Assumption \ref{assume1} holds.

Using \eqref{trailing:eq:41} and $f(x)=(1-\alpha)x$ we obtain that 
\be H(z)=z^{\frac{1-\beta^-}{2\gamma}}-Kz^{\frac{-\epsilon-\beta^-}{2\gamma}},\quad\varphi(z)=(z^{\frac{1}{2\gamma}}(1-\alpha))^{2\gamma}=(1-\alpha)^{2\gamma}z=:\bar{\alpha}z.\label{trailing:eq:43}\ee
\[H''(z)=n_1(n_1-1)z^{n_1-2}-K n_2(n_2-1)z^{n_2-2}=z^{n_2-2}[n_1(n_1-1)z^{n_1-n_2}+K n_2(1-n_2)]\]
\[\half\sigma^2(-\epsilon)(-\epsilon-1)-\mu\epsilon-q<0, \half\sigma^2\epsilon(\epsilon+1)-\mu\epsilon-q<0.\]
It follows that 
\begin{align}
H_f(z)=&\exp(-\int_z^{z_f^\star}\frac{\diff \nu}{\nu-\varphi(\nu)})H(z_f^\star)+\int_z^{z_f^\star}H(\varphi(\nu))\exp(-\int_{z}^{\nu}\frac{\diff w}{w-\varphi(w)})\frac{\diff \nu}{\nu-\varphi(\nu)}\nn\\
=&\bigg(\frac{z}{z_f^\star}\bigg)^{\frac{1}{1-\bar{\alpha}}}[(z_f^\star)^{\frac{1-\beta^-}{2\gamma}}-K(z_f^\star)^{\frac{-\epsilon-\beta^-}{2\gamma}}]+\frac{(\bar{\alpha})^{\frac{1-\beta^-}{2\gamma}}}{(1-\bar{\alpha})\frac{1-\beta^-}{2\gamma}-1}[(z_f^\star)^{\frac{1-\beta^-}{2\gamma}}\bigg(\frac{z}{z_f^\star}\bigg)^{\frac{1}{1-\bar{\alpha}}}-z^{\frac{1-\beta^-}{2\gamma}}]\nn\\
&-K\frac{(\bar{\alpha})^{\frac{-\epsilon-\beta^-}{2\gamma}}}{(1-\bar{\alpha})\frac{-\epsilon-\beta^-}{2\gamma}-1}[(z_f^\star)^{\frac{-\epsilon-\beta^-}{2\gamma}}\bigg(\frac{z}{z_f^\star}\bigg)^{\frac{1}{1-\bar{\alpha}}}-z^{\frac{-\epsilon-\beta^-}{2\gamma}}].\label{trailing:eq:44}
\end{align}
Notice that \eqref{trailing:eq:42} ensures that two detonators in the last line of \eqref{trailing:eq:44} are negative. 

Using \eqref{trailing:eq:41}, \eqref{trailing:eq:43} and \eqref{trailing:eq:44}, we obtain 
\begin{align}
&H^{(1)}(z)\nn\\
=&H_f(z)-H(z)\nn\\
=&\bigg(\frac{(\bar{\alpha})^{\frac{1-\beta^-}{2\gamma}}+(1-\bar{\alpha})\frac{1-\beta^-}{2\gamma}-1}{(1-\bar{\alpha})\frac{1-\beta^-}{2\gamma}-1}(z_f^\star)^{\frac{1-\beta^-}{2\gamma}}-K\frac{(\bar{\alpha})^{\frac{-\epsilon-\beta^-}{2\gamma}}+(1-\bar{\alpha})\frac{-\epsilon-\beta^-}{2\gamma}-1}{(1-\bar{\alpha})\frac{-\epsilon-\beta^-}{2\gamma}-1}(z_f^\star)^{\frac{-\beta^-}{2\gamma}}\bigg)\bigg(\frac{z}{z_f^\star}\bigg)^{\frac{1}{1-\bar{\alpha}}}\nn\\
&-\frac{(\bar{\alpha})^{\frac{1-\beta^-}{2\gamma}}+(1-\bar{\alpha})\frac{1-\beta^-}{2\gamma}-1}{(1-\bar{\alpha})\frac{1-\beta^-}{2\gamma}-1}(z_f^\star)^{\frac{1-\beta^-}{2\gamma}}\bigg(\frac{z}{z_f^\star}\bigg)^{\frac{1-\beta^-}{2\gamma}}+K\frac{(\bar{\alpha})^{\frac{-\epsilon-\beta^-}{2\gamma}}+(1-\bar{\alpha})\frac{-\epsilon-\beta^-}{2\gamma}-1}{(1-\bar{\alpha})\frac{-\epsilon-\beta^-}{2\gamma}-1}(z_f^\star)^{\frac{-\epsilon-\beta^-}{2\gamma}}\bigg(\frac{z}{z_f^\star}\bigg)^{\frac{-\epsilon-\beta^-}{2\gamma}}\nn\\
=:& k(\frac{z}{z_f^\star}),
\end{align}
where $k(u)$ is a polynomial in $u$:
\be
k(u)=Au^{n_1}+Bu^{n_2}+Cu^{n_3},\quad\forall u\in(0,1] \label{kuu}
\ee
with
\[\frac{1}{1-\bar{\alpha}}\equiv n_1>1>n_2\equiv\frac{1-\beta^-}{2\gamma}>n_3\equiv\frac{-\epsilon-\beta^-}{2\gamma}>0,\]
and unambiguous definitions of the coefficients $A,B,$ and $C$. We can show that  $C>0$. In view of the fraction inside $C$, we let $g(x)=x^p+p(1-x)-1$ for $p=\frac{-\epsilon-\beta^-}{2\gamma}\in(0,1)$. Then $g(1)=0$ and $g'(x)=p(x^{p-1}-1)>0$ for all $x\in(0,1)$ so $g(\cdot)$ is strictly increasing over $(0,1)$. In particular, $g(\bar{\alpha})<g(1)=0$. Since the denominator in $C$ is also negative, we conclude that $C>0$.

Also, observe that  $k(0+)=0=H^{(1)}(0+)$.  Now, taking derivative of $k(u)$ in \eqref{kuu}, we get
\be
u^{1-n_3}k'(u)=An_1u^{n_1-n_3}+Bn_2u^{n_2-n_3}+Cn_3.
\ee
From $\lim_{u\downarrow 0}u^{1-n_3}k'(u)=Cn_3>0$ we know that $H^{(1),\prime}(z)>0$ for sufficiently small $z>0$. Moreover,
\be u^{2-n_3}k''(u)=An_1(n_1-1)u^{n_1-n_3}+Bn_2(n_2-1)u^{n_2-n_3}+Cn_3(n_3-1). \label{trailing:eq:47}\ee
Using standard argument by taking the derivative, it can be shown that functions like the right hand side of \eqref{trailing:eq:47} can change monotonicity at most once over $(0,1)$. Clearly, the  right hand side of \eqref{trailing:eq:47} converges to $Cn_3(n_3-1)<0$ as $u\downarrow0$. On the other hand, because $H_f(z)-H(z)$ is convex over $(\psi_q(x_0), \psi_q(b_f^\star))$  (see Lemma \ref{trailing:lem:convex}), we know that the  right hand side of \eqref{trailing:eq:47} is positive as $u\uparrow 1$.  Given that $k(u)$ is maximized at $\frac{\ol{z}_f^\star}{z_f^\star}$, we know that  $k''(u)$ changes sign exactly once over $(0,1)$. More specifically, there are $u_1\in(0,1)$ such that  $k''(u)<0$ for all $u\in(0,u_1)$, and $k''(u)>0$ for all $u\in(u_1,1)$. This proves the pattern of convexity change for $H^{(1)}(\cdot)$. It follows that $H^{(1)}(\cdot)$ is strictly increasing from $0$  to $\ol{z}_f^\star$,  in particular, $H^{(1)}(z)>0$ for all $z\in(0,z_f^\star)$. 
 Thus, the smallest nonnegative concave majorant of $H^{(1)}(\cdot)$ is given by 
\[H^{(1)}(z\wedge \ol{z}_f^\star),\]
and the optimal stopping region for \eqref{eq:compound2} is given by $\psi_q^{-1}((0,\ol{z}_f^\star])=(0,\ul{b}_f^\star]$. Finally, the global maximum $\ol{z}_f^\star$ is the unique solution to 
\be
\label{eq:trailing:olz} k'(\frac{z}{z_f^\star})=0, \frac{\ol{z}_f^\star}{z_f^\star}\in(0,u_1).
\ee
\end{proof}

\bibliographystyle{apa}
\bibliography{ospbib2}   
\end{document}